\documentclass[showpacs,aps,twocolumn, nofootinbib]{revtex4-1}
\usepackage{bm}
\usepackage{amsmath}
\usepackage{graphicx}
\usepackage{subfigure}
\usepackage[usenames,dvipsnames]{color}
\definecolor{darkblue}{RGB}{0,0,196}
\usepackage{setspace}
\usepackage{hyperref}
\usepackage{xcolor}
\usepackage{xspace}
\usepackage{tikz}
\usetikzlibrary{arrows.meta}
\hypersetup{ 
  colorlinks   = true, 
  urlcolor     = blue, 
  linkcolor    = blue, 
  citecolor   = blue 
}
\usepackage{footmisc}
\usepackage[makeroom]{cancel}
\usepackage{comment}
\def\be{\begin{equation}}
\def\ee{\end{equation}}
\def\ba{\begin{eqnarray}}
\def\ea{\end{eqnarray}}
\usepackage{graphicx}
\usepackage{amsmath,bbm}
\usepackage{amssymb,bm}
\usepackage{yfonts}

\newcommand{\pt}{\ensuremath{p_{\rm T}}\xspace}

\begin{document}
\title{Evolution of the underlying event and non-extensive thermodynamics \\ in pp collisions from RHIC to LHC energies}
\author{Suraj Prasad$^{1}$}\email[]{suraj.prasad@cern.ch}
\author{Aditya Nath Mishra$^{2}$}\email[]{adityanathmishraphy@gmail.com}
\author{Guy Pai{\'c}$^{1,3}$}\email[]{guy.paic@cern.ch}
\author{Gergely G\'abor Barnaf\"oldi{$^{1}$}}\email[]{barnafoldi.gergely@wigner.hun-ren.hu}
\affiliation{$^{1}$HUN-REN Wigner Research Centre for Physics, 29-33 Konkoly-Thege Mikl\'os Road, 1121 Budapest, Hungary}
\affiliation{$^{2}$C. M. Science College, L. N. Mithila University, Darbhanga, India}
\affiliation{$^{3}$Instituto de Ciencias Nucleares, Universidad Nacional Aut\'onoma de M\'exico, \\ Apartado Postal 70-543, M\'exico City 04510, M\'exico}

\begin{abstract}
We investigate the geometrical and thermodynamical aspects of particle production in pp collisions over a broad center-of-mass energy, ranging from $\sqrt{s}=62.4$ GeV to 13 TeV using PYTHIA8. We study the azimuthal dependence of charged-particle production relative to the leading particle. This includes the study of the Tsallis--Pareto parameters extracted from the charged-particle transverse-momentum spectra in fixed $\Delta\phi$ intervals relative to the leading charged particle. The simultaneous analysis of charged-particle multiplicity and event topology demonstrates that the underlying event (UE) dominated region can be extended beyond the conventional transverse side region across RHIC and LHC energies. To further disentangle effects from soft and hard particle production mechanisms, the study is performed in different transverse spherocity and charged-particle flattenicity classes. We demonstrate that the UE region extends beyond the conventional transverse region, validating an enlarged angular interval of $40^\circ\lesssim|\Delta\phi|\lesssim140^\circ$ from RHIC to LHC energies. Event-shape selections reveal that jetty events consistently exhibit larger $q$ and lower $T_{\rm s}$ than isotropic events. The $\Delta\phi$ profiles of charged-particle multiplicity and transverse spherocity remain universal in shape across collision energies despite the strong increase in overall event activity, whereas the Tsallis parameters retain a residual, leading-particle-driven energy dependence in the near- and away-side regions.
\pacs{}
\end{abstract}
\date{\today}
\maketitle 

\section{Introduction}
\label{intro}
Measurements at the Relativistic Heavy-Ion Collider (RHIC) and the Large Hadron Collider (LHC) are widely interpreted as evidence that high-energy nucleus--nucleus collisions possibly produce a strongly interacting medium, commonly referred to as the quark--gluon plasma (QGP), whose collective behavior is often described phenomenologically in terms of near-perfect-fluid hydrodynamics~\cite{Arsene:2005,BBBack:2005,Adams:2005,Adcox:2005,Chatrchyan:2012,Abelev:2013,Aad:2013}.
Traditionally, proton--proton (pp) collisions have been regarded as a benchmark for perturbative quantum chromodynamics (pQCD) processes, providing a baseline for identifying nuclear modifications. Proton (or deuteron)--nucleus collisions have been used to study initial- and final-state effects in the absence of QGP formation.
However, recent measurements at the LHC have revealed features in pp and p--Pb collisions with high final-state multiplicity that are reminiscent of signatures previously considered unique to QGP formation~\cite{CMS:2010ifv, ATLAS:2015hzw, CMS:2015fgy, CMS:2016fnw, CMS:2012qk, ALICE:2012eyl}. To date, it remains an open question whether QGP droplets can form in such small systems, or if these observed features instead arise from complex vacuum QCD effects~\cite{CMS:2026qef, ALICE:2026zck, ATLAS:2026fos, ATLAS:2026zgq}.

In this regard, the study of hadron production in pp collisions provides important insight into the perturbative and non-perturbative regime of QCD. While hard-scattering processes, which are dominant in the high-$p_{\rm T}$ regime, can be described using pQCD; the majority of produced particles originate from soft-QCD processes including multiparton interactions (MPI), color reconnection, beam remnants, and hadronization mechanisms that remain only partially understood due to asymptotic nature of QCD. These soft-QCD processes could be linked to the majority of the enhanced particle-production features observed in high-multiplicity pp collisions~\cite{CMS:2016fnw, Mishra:PRC2019}. In high-energy collisions, these soft contributions form the Underlying Event (UE) activity, which accompanies the primary hard scattering and significantly influences the final-state event topology and particle production characteristics~\cite{Field:2000dy, CDFPaper2, Agocs}.

The non-extensive statistical approaches based on Tsallis thermodynamics have proven particularly successful in describing transverse-momentum ($p_{\rm T}$) spectra over a broad momentum range in hadronic and nuclear collisions~\cite{Tsallis:1987eu, Tsallis:2009tri, Biro:2004qg, tsbiro:pa2015, wong}. The Tsallis--Pareto distribution provides a unified description of low-$p_{\rm T}$ thermal-like particle production and high-$p_{\rm T}$ power-law tails associated with perturbative scatterings. Moreover, it successfully describes particle production from small to large colliding systems, for hadrons containing light as well as heavy quarks~\cite{Gyulai:2024qov,  Gyulai:2024dkq, Biro:2020kve,Gyulai:2024dkq,Gemes:2020,Vertesi:2020}. Within this framework, the Tsallis temperature parameter ($T_{\rm s}$), and the non-extensivity parameter ($q$), encode important information about the deviation from equilibrium-like (Boltzmann--Gibbs) behavior, and fluctuations of the produced system. Tsallis parameters are sensitive to system size and event shape, making them useful probes of the underlying dynamics of particle production~\cite{biro, Varga:2019, Shen:2019, Shen:2017}.

In a recent study, the angular dependence of multiplicity distributions, transverse-momentum spectra, and event-shape observables was investigated in pp collisions at $\sqrt{s}=13$ TeV using PYTHIA8~\cite{Mishra:2021hnr}. A sliding-angle-based angular selection scheme was introduced there to characterize the event structure, which explores the geometrical structure of the UE. The sliding-angle approach is of particular interest, which enabled a differential mapping of the azimuthal structure of the event using fixed-width $\Delta\phi$ intervals relative to the leading (or highest $p_{\rm T}$) charged particle. 
The study revealed that the traditional UE definition can be extended beyond the conventional transverse side (TS) region and demonstrated that the Tsallis parameters attain remarkably stable values within this UE-dominated angular domain. Furthermore, at $\sqrt{s}=13$~TeV the UE region was found to lie closer to the Boltzmann--Gibbs limit, $q \rightarrow 1$, than the jet-dominated regions, suggesting a more isotropic and weakly non-extensive environment. Although these observations provided important insight into the geometrical and thermodynamical structure of the UE at LHC energies, it remains unclear whether these identified patterns across observables are universal across collision energies or emerge only at sufficiently high $\sqrt{s}$, where MPI activity becomes dominant. Since the charged-particle multiplicity density, the probability of multiple partonic interactions, and the hardness of the spectra all evolve strongly with collision energy, a systematic energy-dependent investigation is necessary to establish the scaling behavior and universality of the Tsallis parameters and UE topology.

Moreover, the study is extended using different event classifiers, namely, transverse spherocity~\cite{ALICE:2023bga, Cuautle:2014yda, Ortiz:2015ttf, Prasad:2025yfj, Prasad:2024gqq} and charged-particle flattenicity~\cite{Ortiz:2022mfv, Ortiz:2022zqr, ALICE:2024vaf,Prasad:2025yfj, Prasad:2024gqq}. While transverse spherocity has the event classification capability based on the azimuthal distribution of particles and characterizes the transverse-momentum distribution of the event, charged-particle flattenicity considers the particle distributions along longitudinal and azimuthal directions and emphasizes the uniformity of particle distribution. Therefore, these two classifiers are sensitive to different aspects of the event topology and together can provide a more differential understanding of soft and hard QCD processes~\cite{Prasad:2025yfj, Prasad:2024gqq}.

Event-shape observables and thermodynamical quantities probe complementary aspects of particle production in high-energy hadronic collisions. While spherocity and flattenicity characterize the geometrical emission of particles in momentum space, the Tsallis parameters quantify the degree of non-extensivity, fluctuations, and information content of the produced system. 

In this work, we study proton--proton collisions over a broad center-of-mass energy range, $\sqrt{s}=62.4$ GeV to 13 TeV, generated using PYTHIA8. Although Tsallis thermodynamics, event-shape observables, and underlying-event studies have each been investigated separately in previous works, their simultaneous, energy-dependent characterization in localized azimuthal regions across RHIC to LHC energies has not yet been systematically explored. The present work addresses this gap by combining these complementary approaches within a unified multi-differential framework. We investigate the energy evolution of Tsallis--Pareto parameters extracted from charged particle $p_{\rm T}$-spectra in fixed azimuthal intervals relative to the leading charged particle. Furthermore, the study is performed in different spherocity and flattenicity classes to disentangle the contributions of jet-dominated and isotropic particle-production mechanisms. The objective is to explore possible universal scaling properties of the UE and to establish connections between event geometry, 
and UE dynamics from RHIC to LHC energies. Such a multidimensional approach provides a unique opportunity to quantify how the geometrical and thermodynamical characteristics of particle production evolve with collision energy and event topology, thereby offering new constraints on the mechanisms responsible for the observed particle-production phenomenology in small collision systems.

The rest of the paper is organized as follows. Section~\ref{pythia} introduces the event generation settings, Tsallis formalism, event-shape definitions, and the sliding-angle analysis. Section~\ref{results} presents the energy and event-topology dependence of multiplicity, Tsallis parameters. 
Finally, we summarize our findings and provide an outlook for future measurements and model constraints in Sec.~\ref{summary}.

\section{Event Generation and methodology}
\label{pythia}

\subsection{The model: PYTHIA8}
PYTHIA8~\cite{pythia8.2} is a versatile Monte Carlo event generator commonly used to simulate events in high-energy leptonic, hadronic and nuclear collisions. It has been refined over several decades through continuous development and tuning against collider and other experimental data. The event generation process in PYTHIA8 typically begins with a hard scattering event, followed by initial- and final-state parton showers, MPI, color reconnection, hadronization, and finally the simulation of hadron decays. The Monash 2013 tune is established to reproduce a wide set of observables from LHC data, while the energy dependence was determined using LEP, SPS and Tevatron results~\cite{Skands:2014pea}. 

For this study, we generate approximately a billion minimum-bias events using PYTHIA8 (version 8.317) for the center-of-mass energies: 62.4 GeV, 200 GeV, 900 GeV, 2.76 TeV, 5.02 TeV, 7 TeV and 13 TeV. We have used Monash 2013 tune~\cite{Skands:2014pea} (\texttt{Tune:pp = 14}) with soft QCD processes (\texttt{SoftQCD:inelastic = on}) which provides a sound description of particle $p_{\rm T}$ and $\eta$ spectra~\cite{alice:prl2010, ALICE:vz}. This can be explicitly observed in Fig.~\ref{fig:sqrtsvsnch} (see Appendix~\ref{sec:sqrtsvsnch}), where PYTHIA8 Monash describes the $\sqrt{s}$ dependence of ${\rm d}N_{\rm ch}/{\rm d}\eta$ at $|\eta| < 0.5$ measured by ALICE~\cite{ALICE:2015olq,ALICE:2010cin,ALICE:2015qqj}. The results are presented for the inelastic (INEL), inelastic events with at least one charged particle in the acceptance (INEL$>0$), and non-single diffractive (NSD) event classes. The agreement between PYTHIA8 and the experimental measurements over a broad $\sqrt{s}$ range demonstrates the suitability of PYTHIA8 for the present study.

The event and particle selection criteria are defined as follows. Events are required to contain at least three primary charged particles with transverse momentum $\pt{}>0.15$ GeV/$c$ within the pseudorapidity range $|\eta|<0.8$.  This condition is imposed to ensure compatibility with the later-defined event-shape variable calculations. Additionally, events are required to have at least one charged particle in the pseudorapidity region $-3.4<\eta<-2.3$ or $3.8<\eta<5.0$, which corresponds to the FT0 detector coverage at ALICE.
Primary charged particles are defined as final-state charged particles, including decay products, except those originating from weak decays of strange particles. This definition aligns closely with the one adopted by the ALICE experiment~\cite{alice:2017}, making it suitable for potential future experimental comparisons. It should be noted that, because the charged-particle density decreases strongly with decreasing $\sqrt{s}$, the above requirements retain a significantly smaller fraction of inelastic events at RHIC than at LHC energies. This behavior at the lowest energies should be kept in mind when interpreting the RHIC-energy results discussed in Sec.~\ref{sec:delphivsTandq}.

\subsection{Tsallis formalism}
\label{sec:tsallissec}
The transverse momentum ($p_{\rm T}$) spectra of particles produced in high-energy collisions receive contributions from both soft and hard processes. While the low-$p_{\rm T}$ region is dominated by soft particle production, the high-$p_{\rm T}$ region is increasingly influenced by hard partonic scatterings. The Tsallis formalism has been widely used to describe the entire $p_{\rm T}$ spectrum within a single statistical framework by incorporating possible deviations from the conventional Boltzmann--Gibbs equilibrium through a non-extensive parameter, $q$.

In the present work, the charged-particle $p_{\rm T}$ spectra are fitted using the thermodynamically consistent Tsallis--Pareto distribution~\cite{sorin,Azmi:2014dwa,biro,parvan2,azmi, parvan2006},
\begin{equation}
\frac{1}{2\pi p_{\rm T}}\frac{\mathrm{d}^{2}N}{\mathrm{d}p_{\rm T}\mathrm{d}\eta}
=
A\,m_{\rm T}
\left[
1+(q-1)\frac{m_{\rm T}}{T_{\rm s}}
\right]^{-\frac{q}{q-1}},
\label{eq:TsallisFit}
\end{equation}
where $A$ is the normalization constant, $q$ is the non-extensive parameter, $T_{\rm s}$ is the Tsallis temperature, $m$ is the particle mass, and $m_{\rm T}=\sqrt{p_{\rm T}^{2}+m^{2}}$ is the transverse mass. The parameter $T_{\rm s}$ The parameter $T_{\rm s}$ sets the slope of the exponential (soft) component and therefore primarily determines the spectral shape in the low- and intermediate-$p_{\rm T}$ regions, whereas $q$ controls the high-$p_{\rm T}$ power-law tail associated with non-equilibrium effects and hard scattering processes.

It is important to emphasize that $q$ and $T_{\rm s}$ are not independent measures of the spectral hardness at intermediate $p_{\rm T}$. At a fixed mean $p_{\rm T}$, these two Tsallis parameters are strongly anti-correlated: an increase in $q$, which transfers spectral weight into the power-law tail, is accompanied by a decrease in $T_{\rm s}$, and vice versa~\cite{Bhattacharyya:2017rlp,Patra:2020qbr}. Therefore, $T_{\rm s}$ alone cannot be used to quantify the hardness of the $p_{\rm T}$-spectra; the pair $(q,,T_{\rm s})$ must be considered together. This does not imply that $q$ and $T_{\rm s}$ should be positively correlated when comparing event classes with different $\langle p_{\rm T}\rangle$, as observed with the Tsallis thermometer in Ref.~\cite{Gyulai:2026}, where $T_{\rm s}$ was found to increase with $\langle p_{\rm T}\rangle$. The two statements refer to different types of comparisons. The anti-correlation is a property of the fit at fixed $\langle p_{\rm T}\rangle$, whereas the simultaneous increase of $q$ and $T_{\rm s}$ discussed in Sec.~\ref{sec:tsallisthermo} arises when comparing event classes with different $\langle p_{\rm T}\rangle$. These two behaviours are therefore used separately in Sec.~\ref{sec:delphivsTandq} and Sec.~\ref{sec:tsallisthermo}.

The spectra are fitted independently for each event class using a $\chi^{2}$ minimization procedure.
The spectra are fitted independently for each event class and each $\Delta\phi$ bin using a $\chi^{2}$ minimization procedure over the range $p_{\rm T}>0.15$~GeV/$c$, assuming the pion mass for the inclusive charged-particle sample.

Equation~\eqref{eq:TsallisFit} follows from the corresponding Tsallis occupation function assuming vanishing chemical potential ($\mu=0$), given as follows:
\begin{equation}
f(E)=
\left[
1+(q-1)\frac{E}{T_{\rm s}}
\right]^{-\frac{1}{q-1}},
\label{eq:occupation}
\end{equation}
where $E=\sqrt{p^{2}+m^{2}}$ is the particle energy. 

Throughout this work, the fitted parameters $A$, $T_{\rm s}$, and $q$ are used to investigate the interplay between soft particle production, hard partonic scatterings, and non-extensive dynamics in proton--proton collisions.

\subsection{Event shape classifiers}
In pQCD-based models, a collision event consists of two microscopic components: a hard scattering, typically dominated by the leading jet, and a soft component, which is primarily associated with the UE activity. In PYTHIA8, a significant fraction of the UE originates from MPIs. Although the number of MPIs ($N_{\rm mpi}$) cannot be measured directly in experiments, several event-shape classifiers have been shown to correlate strongly with the underlying MPI activity~\cite{Banfi:2010xy}. We use two such classifiers, namely, transverse spherocity and charged-particle flattenicity in this study, which are described as follows. Their combined evolution provides complementary information on both the geometrical and thermodynamical characteristics of particle production.

\subsubsection{Transverse spherocity}
The transverse spherocity ($S_{0}$) is defined for a unit vector $\hat{n}$ in the transverse plane which minimizes the ratio~\cite{ALICE:2023bga, Cuautle:2014yda, Ortiz:2015ttf, Prasad:2025yfj}:
\begin{equation}
S_{0}=\frac{\pi^2}{4}\min_{\hat{n}}\Bigg(\frac{\sum_{i=1}^{N_{\rm ch}} |\vec{p}_{{\rm T}, i}\times \hat{n}|}{\sum_{i=1}^{N_{\rm ch}}|\vec{p}_{{\rm T},i}|}\Bigg)^{2},
\label{eq:sphero}
\end{equation}
where, $p_{{\rm T}, i}$ is the transverse momentum of $i$\textsuperscript{th} charged particle and $N_{\rm ch}$ is the total number of charged particles in $|\eta|<0.8$ with $p_{\rm T}>0.15$ GeV/$c$. $\pi^2/4$ is the normalization factor which ensures that $S_0$ lies between 0 and 1. Two extreme limits of $S_0$, i.e., $S_{0}\rightarrow0$ and 1 refer to the jetty and isotropic topology of particle emission in the transverse plane, respectively. We consider only those events for the calculations which have at least 3 charged particles satisfying $|\eta|<0.8$ and $p_{\rm T}>0.15$ GeV/$c$~\cite{Mishra:2021hnr}. Based on the transverse spherocity selection, the 20\% of events with the lowest $S_{0}$ values ($0$--$20\%$ $S_{0}$) are classified as jetty events, whereas the 5\% of events with the highest $S_{0}$ values ($95$--$100\%$ $S_{0}$) are classified as isotropic events. The percentile boundaries are calculated separately for each $\sqrt{s}$, which makes the jetty and isotropic classes always correspond to the same fraction of the selected event sample at different $\sqrt{s}$. 

\subsubsection{Charged-particle flattenicity}
The newly introduced event shape classifier, charged-particle flattenicity ($\rho_{\rm ch}$), is designed to have minimum selection bias, as the calculation requires charged-particle information in forward and backward pseudorapidity regions rather than midrapidity regions, where most of the particle properties are studied. For this study, we calculate $\rho_{\rm ch}$ in 
ALICE FV0 detector acceptance region, i.e., $2.2<\eta<5.0$ where the ($\eta-\phi$) space is divided into ($8\times 8$) cells and charged particles in each cell `$i$' ($N_{\rm ch}^{\rm{cell}, i}$) are estimated. Charged-particle flattenicity is calculated using the following expression~\cite{Prasad:2024gqq, Prasad:2025yfj,Ortiz:2022mfv}.
\begin{equation}
    \rho_{\rm ch} = \frac{\sqrt{\sum_{i}\left(N_{\rm ch}^{\rm{cell}, i}-\langle N_{\rm ch}^{\rm{cell}}\rangle \right)^2/N^{2}_{\rm{cell}}}}{\langle N_{\rm ch}^{\rm{cell}}\rangle}
    \label{eq:flatNch}
\end{equation}
Here, $\langle N_{\rm ch}^{\rm cell}\rangle$ represents the mean number of charged particles per cell. By construction, $\rho_{\rm ch}$ varies between 0 and 1, with $\rho_{\rm ch}\rightarrow 0$ corresponding to isotropic events and $\rho_{\rm ch}\rightarrow 1$ corresponding to jetty events. Events with no charged particle in $2.2<\eta<5.0$, for which $\rho_{\rm ch}$ is undefined, are not considered for the flattenicity-based study. Unlike transverse spherocity, flattenicity is evaluated at the forward rapidity and is therefore expected to be less affected by autocorrelation with midrapidity observables. For consistency with the convention adopted for other event-shape observables, we use $(1-\rho_{\rm ch})$ throughout this manuscript. Under this convention, events with $(1-\rho_{\rm ch})\rightarrow 1$ are predominantly isotropic, while those with $(1-\rho_{\rm ch})\rightarrow 0$ are predominantly jetty in nature. In this study, based on the charged-particle flattenicity selection, the 20\% of events with the lowest $1-\rho_{\rm ch}$ values ($0$--$20\%$ $1-\rho_{\rm ch}$) are classified as jetty events, whereas the 5\% of events with the highest $1-\rho_{\rm ch}$ values ($95$--$100\%$ $1-\rho_{\rm ch}$) are classified as isotropic events.

\subsection{Azimuthal region classification}
Experimentally, the UE is characterized using the azimuthal distribution of particles with respect to the highest transverse momentum ($p_{\rm T}$) particle or leading jet in the event. The event is divided into three azimuthal regions based on the relative angle $\Delta\phi = \phi - \phi_{\rm lead}$:

\begin{figure}
    \centering
    \includegraphics[width=\linewidth]{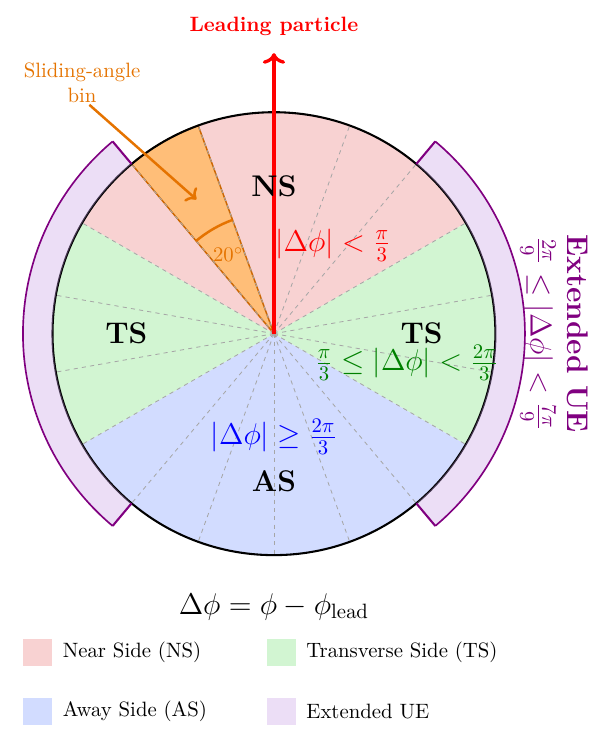}
    \caption{Schematic illustration of the sliding-angle analysis performed with respect to the leading charged particle direction (red arrow). The full azimuthal phase space is partitioned into equal angular intervals of $\Delta\phi=20^\circ$, indicated by the dashed radial lines.}
    \label{fig:slidingAngleCartoon}
\end{figure}

\begin{itemize}
\item {\bf Near Side or NS ($|\Delta\phi| < \frac{\pi}{3}$)}: This region contains the leading particle (or jet) and is therefore dominated by fragmentation of the primary hard scattering.
\item {\bf Away Side or AS ($|\Delta\phi| \geq \frac{2\pi}{3}$)}: Located opposite in azimuth to the leading particle, this region is predominantly populated by the recoil jet from the hard scattering.
\item {\bf Transverse Side or TS ($\frac{\pi}{3} \leq |\Delta\phi| < \frac{2\pi}{3}$)}: Being approximately perpendicular to the hard-scattering axis, this region receives the least contribution from the outgoing hard-scattered partons. Instead, it is particularly sensitive to the UE, including contributions from MPI, beam remnants, and soft initial and final state radiation.
\end{itemize}
Since the NS and AS regions are strongly influenced by hard processes, the transverse region provides the cleanest experimental handle on the UE activity. Measurements of charged-particle multiplicity, scalar $p_{\rm T}$ sum, and particle densities in this region are therefore widely used to quantify the UE and investigate its connection with MPI. As the MPI activity increases, the particle production in the transverse region also increases, making it an effective experimental proxy for the UE.

Ref.~\cite{Mishra:2021hnr} introduced a sliding-angle analysis of leading-particle correlations as a differential probe of the geometrical structure of the event. Instead of partitioning the azimuth into the conventional NS, TS or AS topology, the azimuthal plane was divided into consecutive $20^\circ$ sliding-angle bins. Since each angular interval covers the same geometrical acceptance, this method allows a direct comparison of particle production at different azimuthal locations without biases arising from varying phase-space coverage. By studying the $p_{\rm T}$-spectra, charged-particle multiplicity, and event-shape observables independently in each angular interval, the evolution of soft and hard particle production can be systematically investigated as a function of $\Delta\phi$. This approach revealed that the region associated with the underlying event extends beyond the conventional TS boundaries, proposing an Extended UE region covering $40^\circ \lesssim |\Delta\phi| \lesssim 140^\circ$ ($2\pi/9 \le |\Delta\phi| < 7\pi/9$), corresponding to an approximately $66\%$ larger $\Delta\phi$ coverage than the standard CDF UA definition of the TS. The revised topology, shown in Fig.~\ref{fig:slidingAngleCartoon}, was established from the combined behavior of charged-particle multiplicity and transverse spherocity. Particularly, the region which satisfies the following condition is regarded as the extended UE region.
\begin{equation}
    \bar{S}'_0(\Delta\phi)>\bar{S}_0(\rm MB)
    \label{eq:defExdUE}
\end{equation}
Here $\bar{S}'_0(\Delta\phi)$ and $\bar{S}_0(\rm MB)$ are the event-average transverse spherocity in each bin of $\Delta\phi$ and inclusive $\Delta\phi$ regions, respectively, for minimum bias pp collisions~\cite{Mishra:2021hnr}.
A consequential question is therefore whether the extended UE definition is universal across collision energies or exhibits an energy dependence.
Note that we do not impose any threshold in $p_{\rm T}^{\rm lead}$, unless mentioned otherwise in the figure descriptions. Therefore, the results are interpreted keeping this in mind.

\section{Results}
\label{results}
\subsection{Azimuthal Dependence of Event Topology and Charged-Particle Multiplicity}

A central goal of this paper is to establish how the geometrical and thermodynamical properties of particle production evolve simultaneously as a function of azimuthal angle relative to the leading particle, collision energy, and event topology in pp collisions, which can be achieved within the Tsallis formalism. Before examining the Tsallis parameters, it is informative to characterize the azimuthal structure using observables that are directly sensitive to the hard-process geometry: the normalized charged-particle multiplicity and the average transverse spherocity.

\begin{figure}
	\centering        
	\includegraphics[scale=0.4]{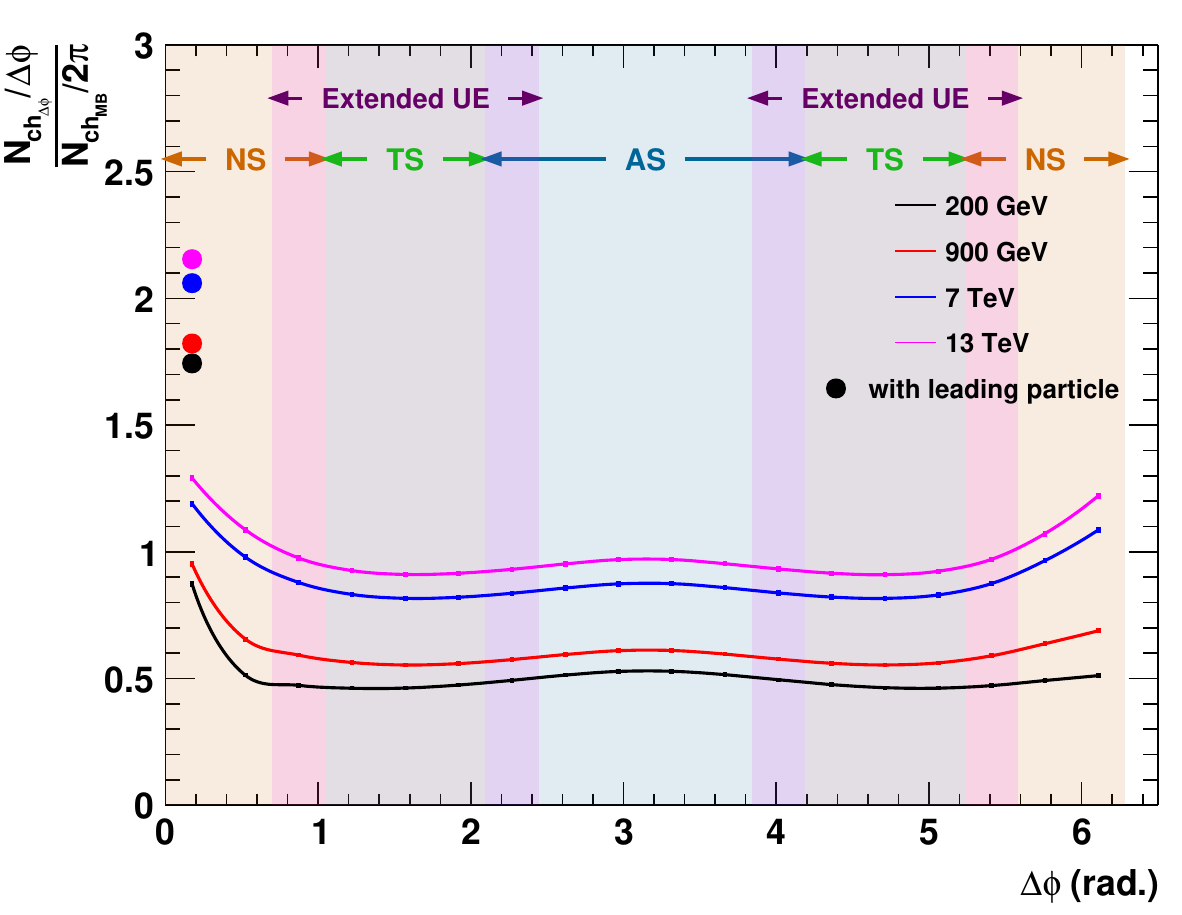}
	\vspace*{-0.5cm}
	\caption{$\Delta\phi$-differential and normalized charged-particle multiplicity for different $\sqrt{s}$ in pp collisions using PYTHIA8 Monash, represented as solid lines. The full markers represent the values with the inclusion of the trigger particle. Here $N_{\rm ch_{\Delta\phi}}$ and $N_{\rm ch_{MB}}$ represent the charged particle multiplicities measured in a $\Delta\phi$ bin, and inclusive region, respectively.}
	\label{fig:NchVsDelPhi}
\end{figure}

Figure~\ref{fig:NchVsDelPhi} shows the $\Delta\phi$-differential normalized charged-particle multiplicity as a function of azimuthal angle relative to the leading charged particle, for center-of-mass energies from $\sqrt{s}=200$ GeV to 13 TeV. The normalized multiplicity as a function of $\Delta\phi$ exhibits the three region structure characteristic of hard-scattering geometry, an enhanced near-side peak centered at $\Delta\phi \approx 0$, an away-side enhancement near $\Delta\phi \approx \pi$, and a relatively flat plateau in the transverse-side region. The near-side and away-side enhancements originate from jet fragmentation of the hard-scattered parton and its recoil particle, respectively. The transverse region which receives minimal contribution from the hard-scattered partons, reflects the UE activity. With an increase in the collision energy, the overall multiplicity rises monotonically, which is driven by the increased MPI probability and higher parton densities at lower Bjorken-$x$, which behavior can be understood by investigating curves from Fig.~\ref{fig:mpiPThatvssqrts} in Appendix~\ref{sec:energydepofmpi}~\cite{Skands:2014pea,ALICE:vz,CDFPaper2,ALICE:2022fnb}. Moreover, the relative hierarchy between different regions is preserved across all $\sqrt{s}$ shown in the figure, confirming that the azimuthal structure of event activity is qualitatively universal from RHIC to LHC energies although the absolute particle yields change significantly. The inclusion of the trigger particle in the near-side (full markers) produces an enhanced value at $\Delta\phi \approx 0$ which indicates the additional contribution of the leading particle.

\begin{figure}
	\centering        
	\includegraphics[scale=0.4]{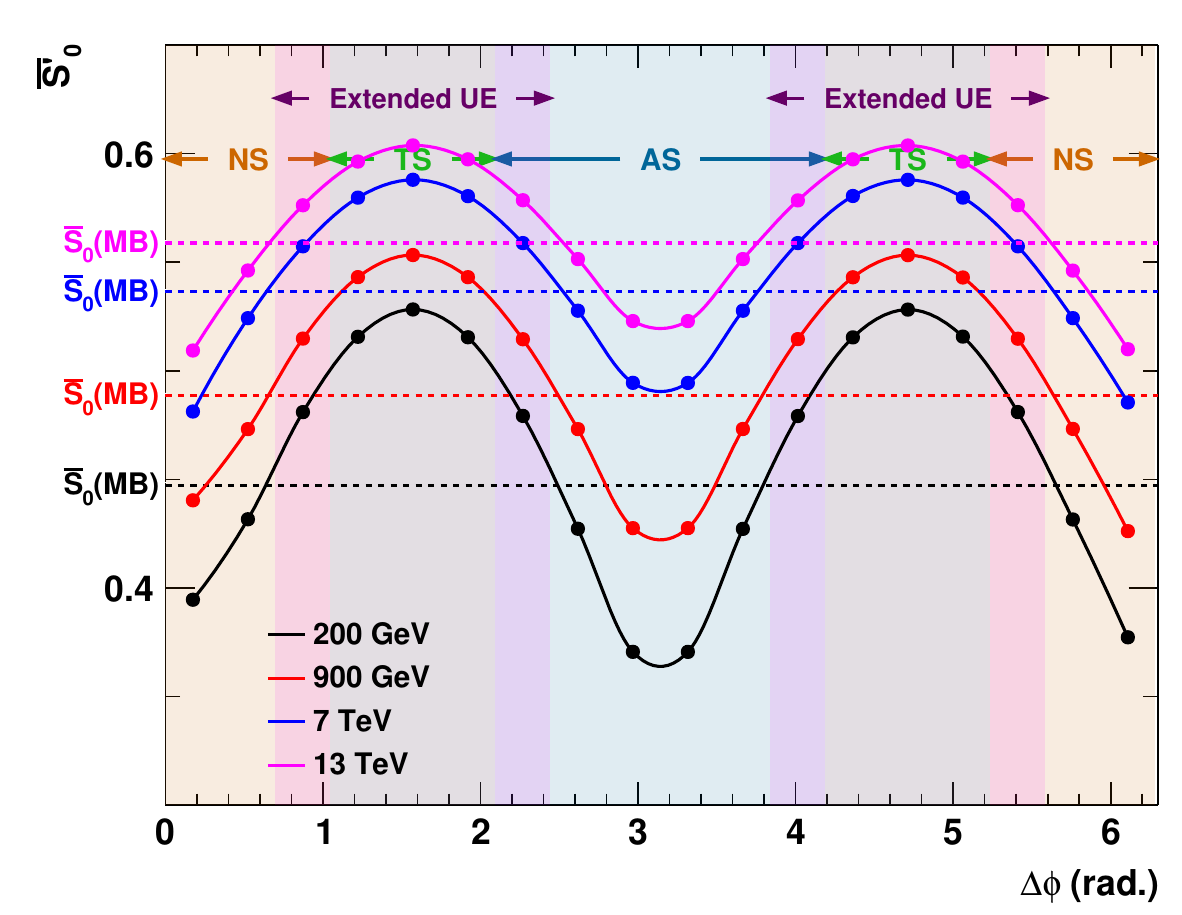}
	\vspace*{-0.5cm}
	\caption{Average transverse spherocity ($\bar{S_{0}^{'}}$) for different sliding angles ($\Delta\phi$) calculated at different $\sqrt{s}$ in minimum-bias pp collisions using PYTHIA8 Monash. The statistical uncertainties are smaller than the marker size.}
	\label{fig:AvgS0VsDelPhi}
\end{figure}

Figure~\ref{fig:AvgS0VsDelPhi} presents the $\Delta\phi$-differential average transverse spherocity, $\bar{S}_{0}^{'}$, for minimum-bias pp collisions from $\sqrt{s}=200$ GeV to 13 TeV using PYTHIA8. The spherocity profile mirrors the multiplicity distribution in a complementary manner. Here, the near-side and away-side regions exhibit a systematically lower $\bar{S}_{0}^{'}$ value, indicating a jetty event topology, while the transverse-side region approaches a higher $\bar{S}_{0}^{'}$, which reflects isotropic emission of particles. This confirms that the transverse region is not merely suppressed in yield but characterized by a more isotropic particle production which is distinct from jet fragmentations. As the collision energy decreases, $\bar{S}_{0}^{'}$ shifts uniformly toward more jet-like values across all azimuthal regions. This behavior is a direct consequence of the reduced MPI activity at lower $\sqrt{s}$: with fewer semi-soft scatterings contributing to the final state, the event topology is increasingly dominated by the leading hard scatter and its fragmentation products~\cite{Skands:2014pea,ALICE:2023bga}.

\begin{figure*}[!t]
	\centering
	\includegraphics[width=0.4\textwidth]{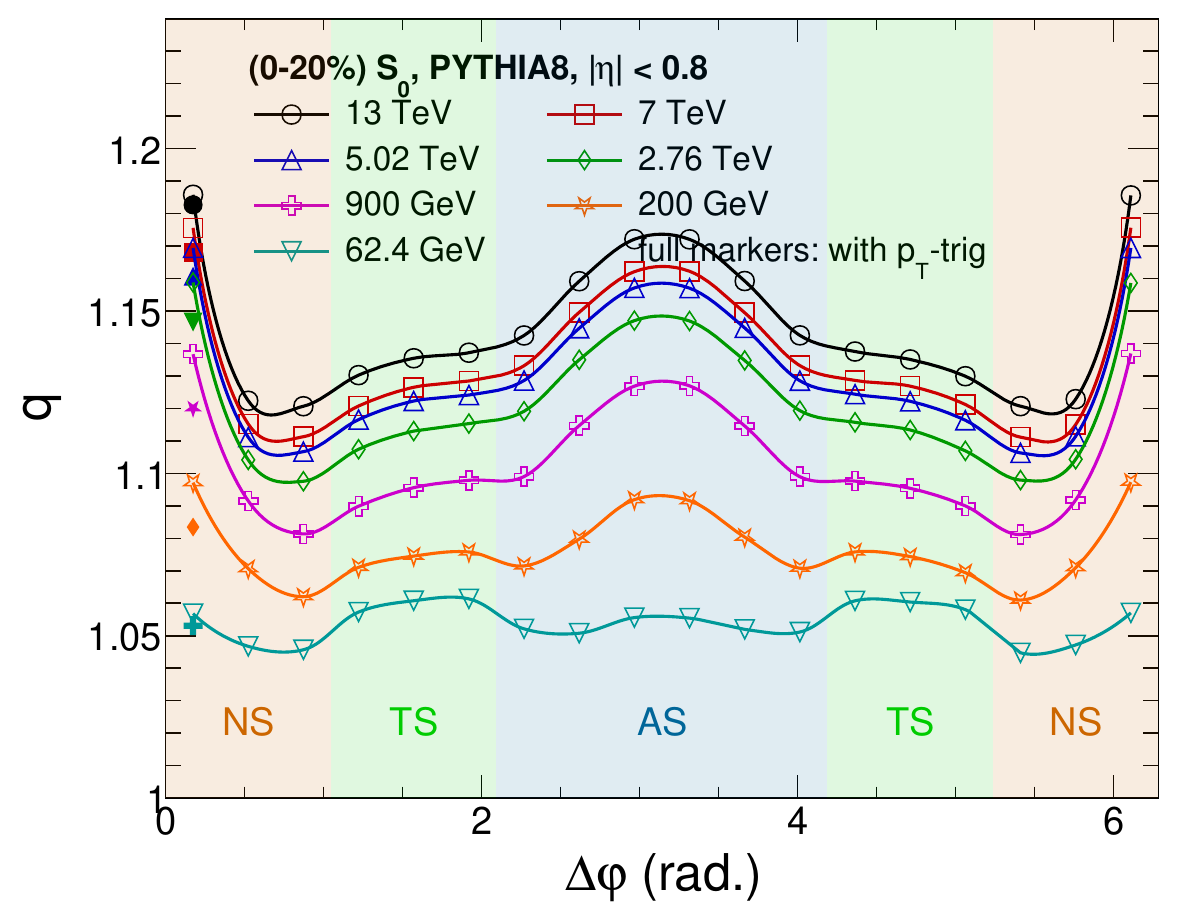}
	\includegraphics[width=0.4\textwidth]{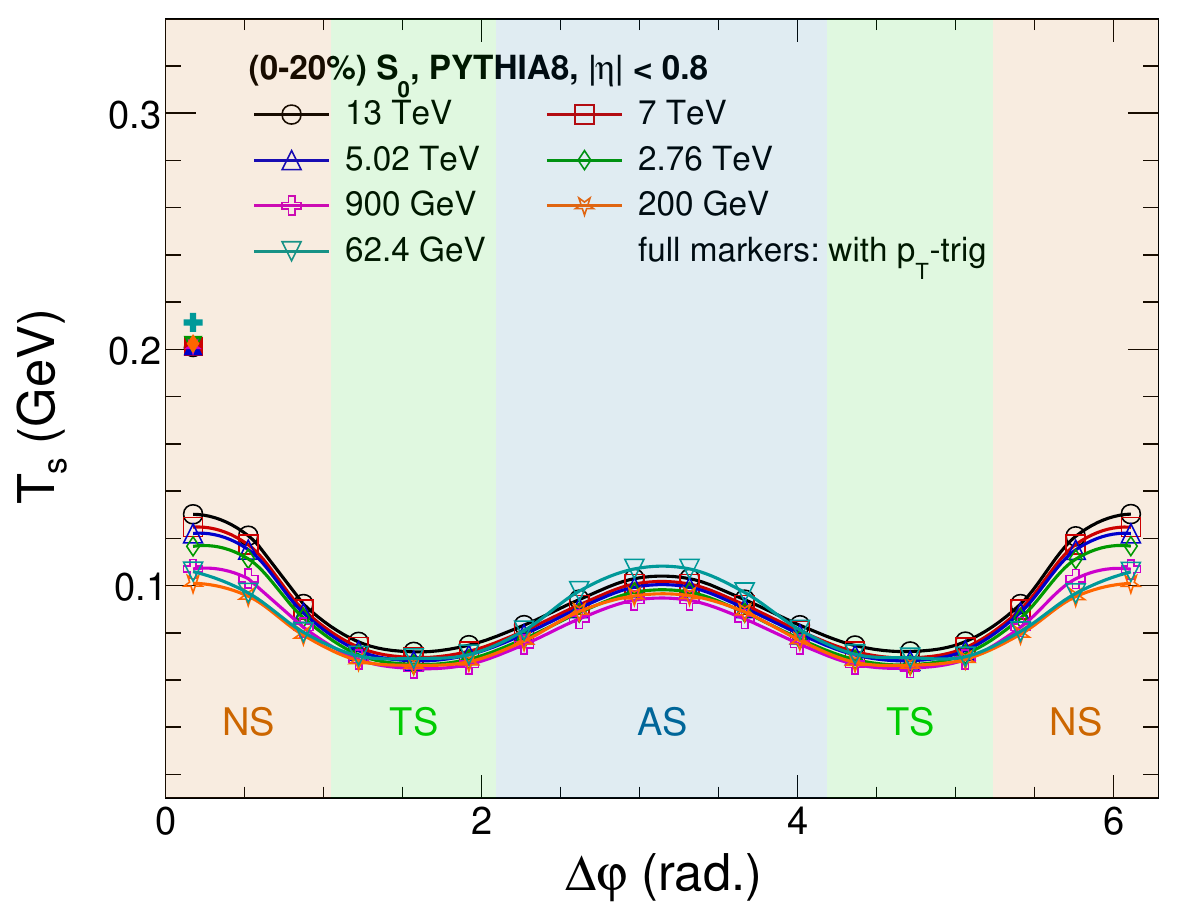}
	\includegraphics[width=0.4\textwidth]{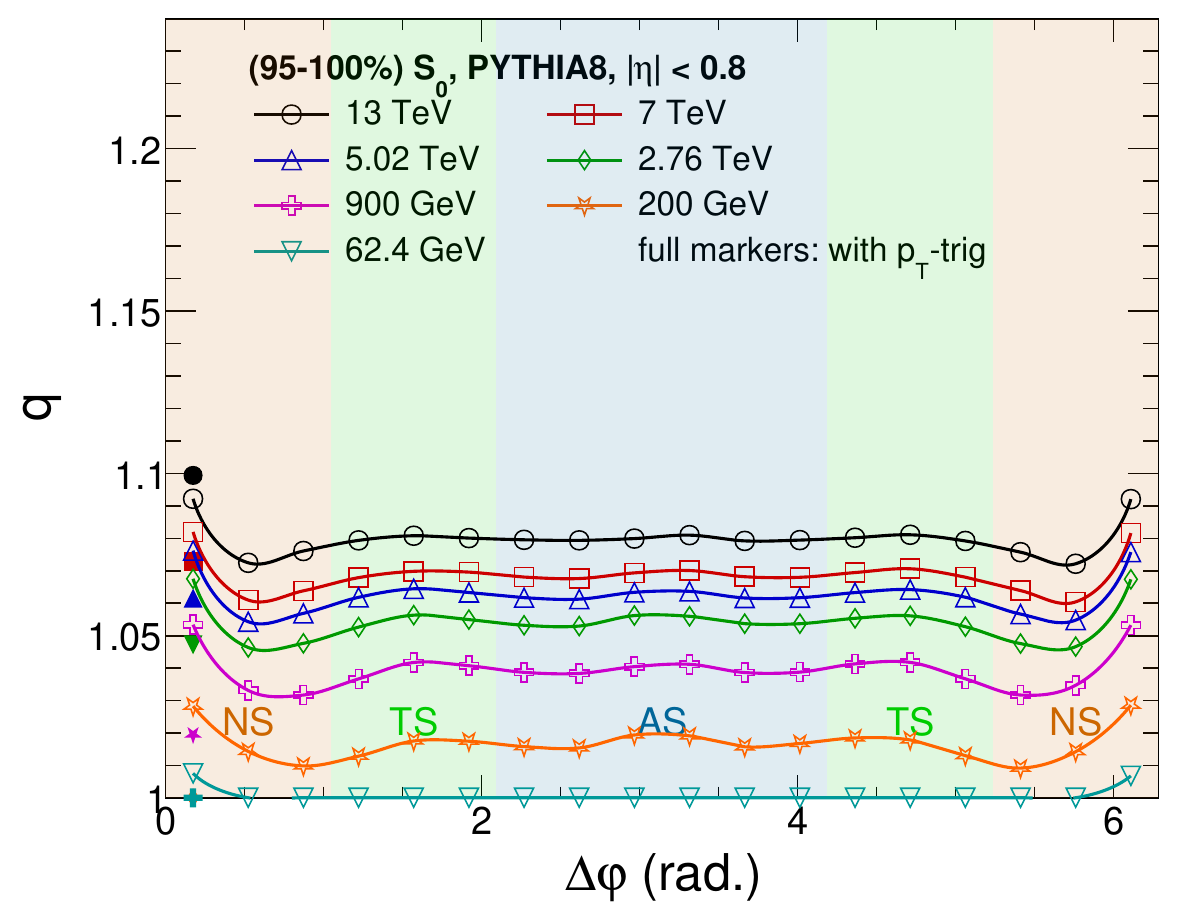}
	\includegraphics[width=0.4\textwidth]{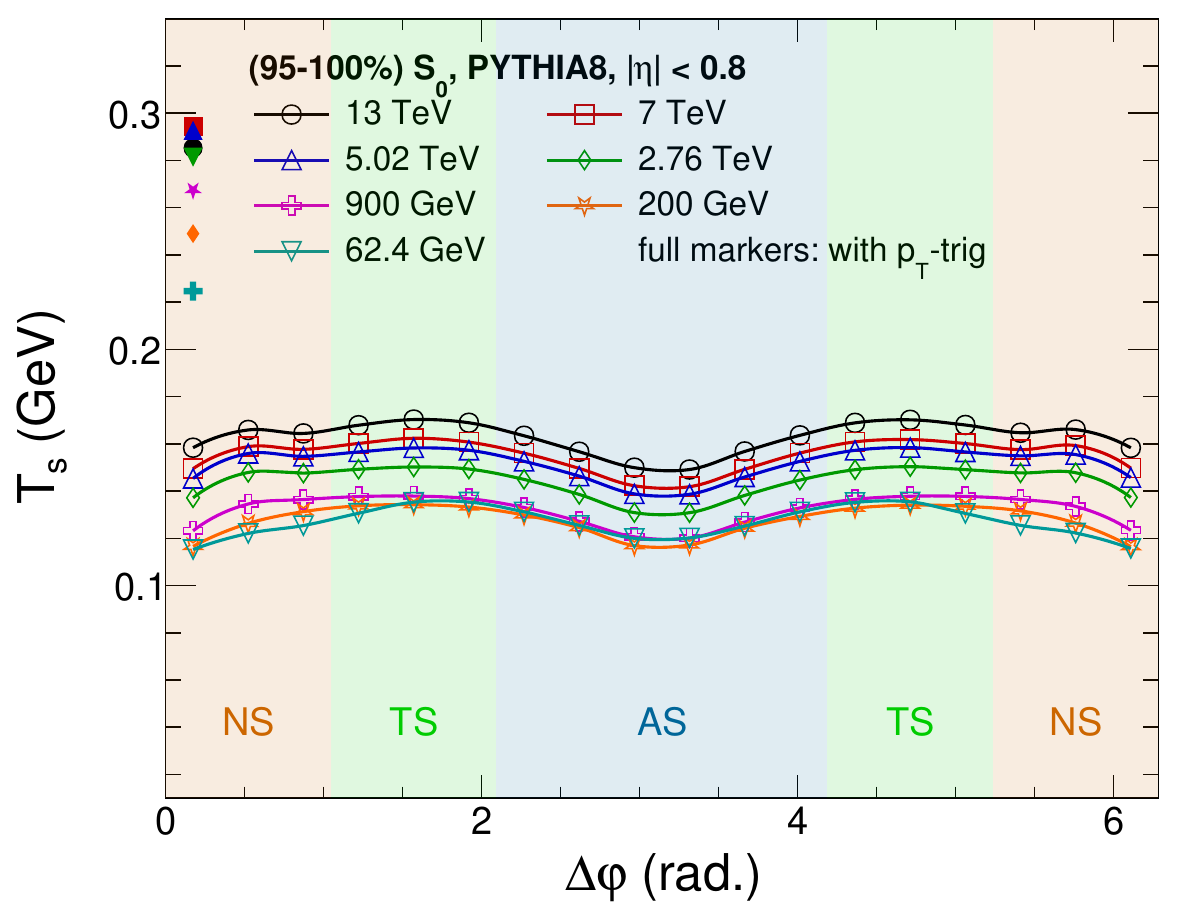}
	\caption{Energy dependence of $q$ (left) and $T_{\rm s}$ (right) as a function of $\Delta\phi$ for two transverse spherocity classes: jetty (upper) and isotropic (lower) pp collisions using PYTHIA8. The statistical uncertainties are smaller than the marker size.}
	\label{fig:s0_qT_energy}
\end{figure*}

An important implication of Figs.~\ref{fig:NchVsDelPhi} and \ref{fig:AvgS0VsDelPhi} is the validation of the extended UE definition, proposed in Ref.~\cite{Mishra:2021hnr}. While the conventional TS region has traditionally been used to characterize the UE, Figs.~\ref{fig:NchVsDelPhi} and \ref{fig:AvgS0VsDelPhi} demonstrate that the particle production and average $S_{0}$ remain nearly unchanged over the wider interval $40^\circ \lesssim |\Delta\phi| \lesssim 140^\circ$ ($2\pi/9 \lesssim |\Delta\phi| \lesssim 7\pi/9$). This indicates that the UE extends well beyond the conventional TS region, making the proposed extended UE definition statistically more robust, owing to its larger angular acceptance, and physically more representative of the UE than the traditional TS definition. In particular the $\bar{S}_{0}^{'}$ remains consistently higher than $\bar{S}_{0}(\rm MB)$ in Fig.~\ref{fig:AvgS0VsDelPhi} indicating soft-QCD dominated particle production with minimal contamination from the NS and AS jet fragments in the extended UE region. Therefore Eq.~\eqref{eq:defExdUE} is satisfied in pp collisions from collision energies ranging from RHIC to LHC.

Figures~\ref{fig:NchVsDelPhi} and \ref{fig:AvgS0VsDelPhi} together convey two key observations. First, the angular decomposition of event activity and topology is qualitatively universal across RHIC and LHC energies. 
Second, although the angular structure is largely energy independent, the overall scales of event activity and topology evolve significantly with $\sqrt{s}$, primarily reflecting the increasing contribution of MPI. These observations naturally motivate the next question: do the thermodynamic properties of the produced particles exhibit a similar universal behavior, and can event-shape selections uncover features that are inaccessible through multiplicity-based classifications alone?

\begin{figure*}[!t]
	\centering        
    \includegraphics[width=0.4\textwidth]{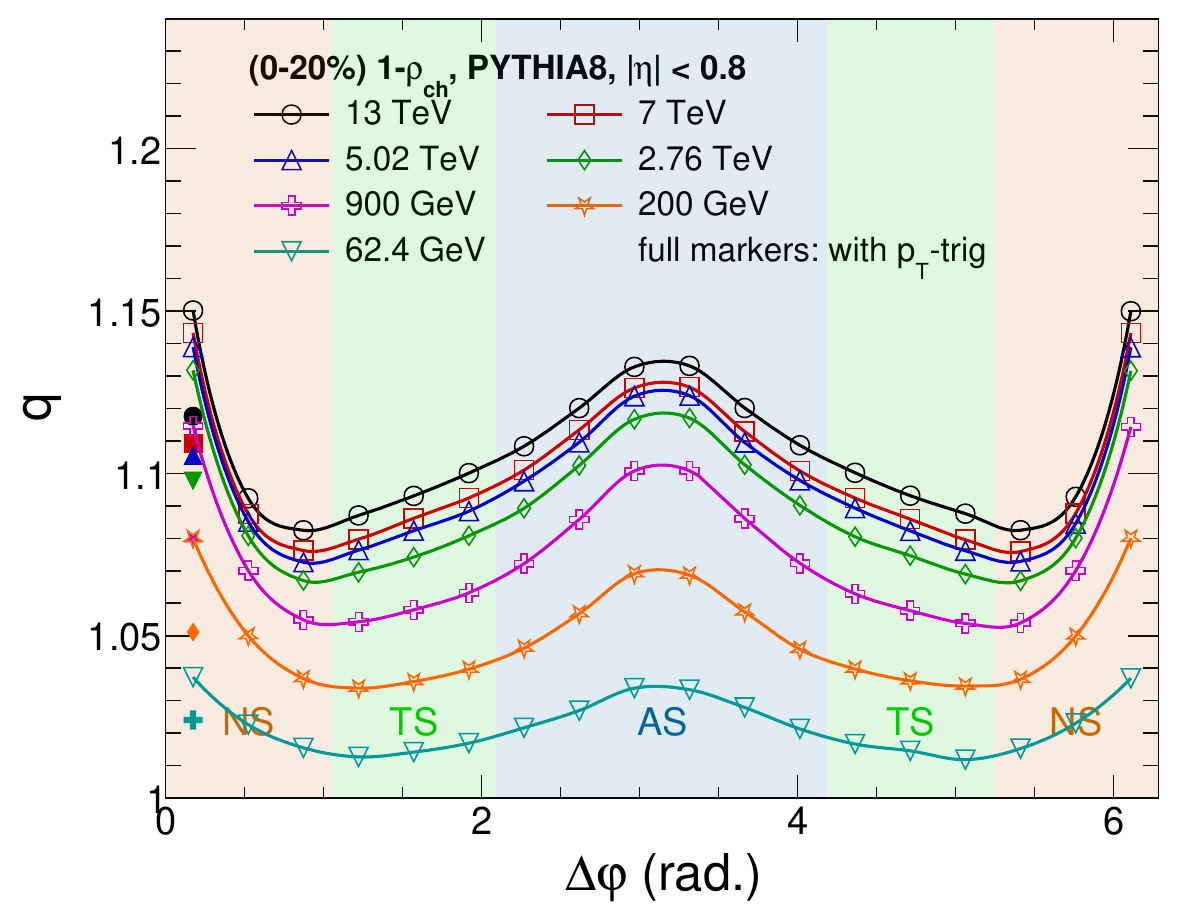}
    \includegraphics[width=0.4\textwidth]{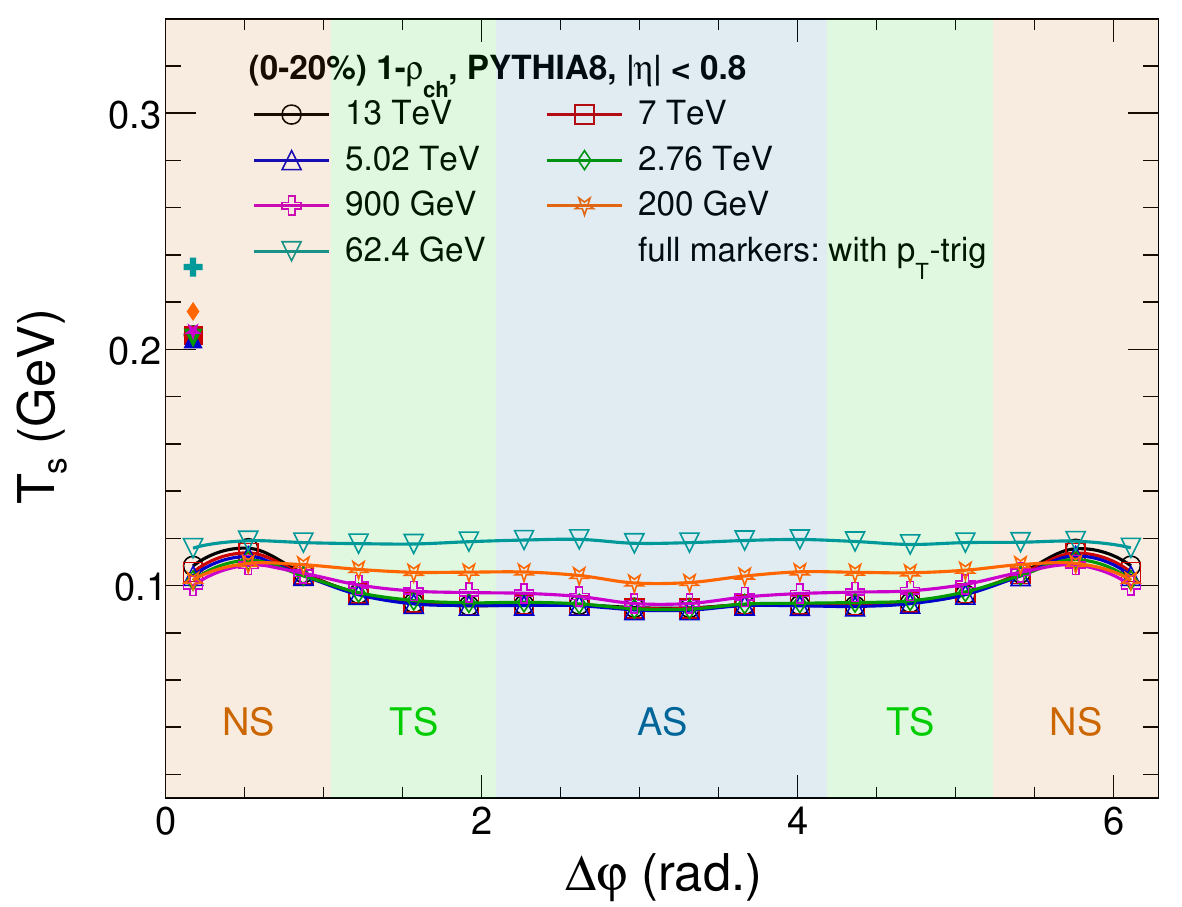}
    \includegraphics[width=0.4\textwidth]{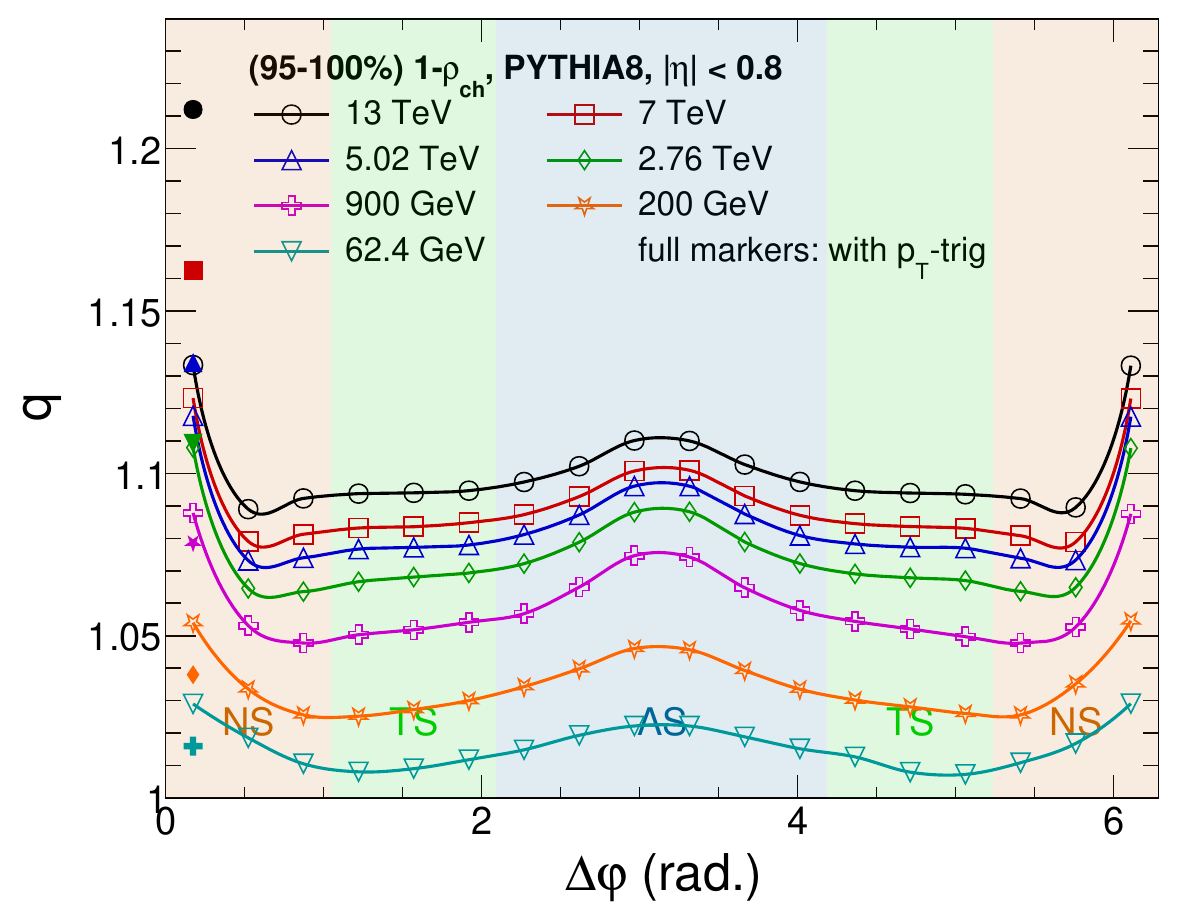}
    \includegraphics[width=0.4\textwidth]{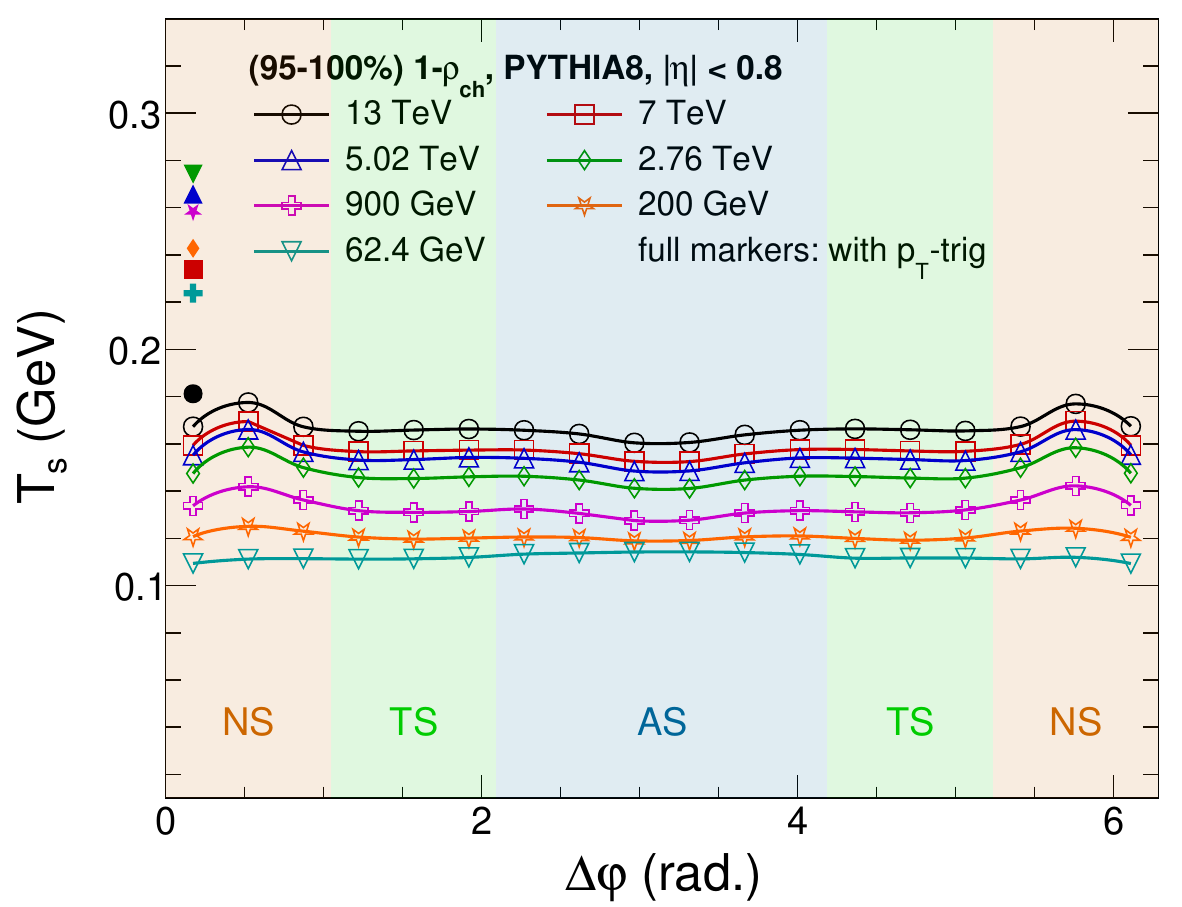}
	\caption{Energy dependence of $q$ (left) and $T_{\rm s}$ (right) as a function of $\Delta\phi$ for two flattenicity classes: jetty (upper) and isotropic (lower) pp collisions using PYTHIA8. The statistical uncertainties are smaller than the marker size.}
	\label{fig:flat_qT_energy}
\end{figure*}

\subsection{Azimuthal Evolution of Tsallis Parameters and Event-Shape Dependence}
\label{sec:delphivsTandq}

The parameter $q$ larger than $1$ shows the excursion of $p_{\rm T}$ spectrum from a pure Boltzmann. 
This non-extensive parameter $q$ also has a statistical meaning in terms of event-by-event fluctuation in the particle multiplicity $n$~\cite{Biro:2020kve}. Modeling the hadrons as a one-dimensional relativistic gas with fluctuating particle number, the non-extensivity parameter appearing in Eq.~\eqref{eq:occupation} can be shown to be fixed by the relative multiplicity fluctuations~\cite{Biro:2020kve},
\begin{equation}
q = 1 - \frac{1}{\langle n \rangle} + \frac{\Delta n^2}{\langle n \rangle^2},
\label{eq:q_multiplicity}
\end{equation}
where $\langle n \rangle$ is the mean event multiplicity and $\Delta n^2 = \langle n^2 \rangle - \langle n \rangle^2$ its variance. In Eq.~\eqref{eq:q_multiplicity}, the three terms have distinct roles: the leading $1$ is the Boltzmann--Gibbs reference value; the term $-1/\langle n \rangle$ is a finite-multiplicity correction that survives even for pure Poissonian statistics and vanishes as $\langle n \rangle \to \infty$; and the term $\Delta n^2/\langle n \rangle^2$ is the non-extensive contribution, capturing multiplicity fluctuations. Equation~\eqref{eq:q_multiplicity} shows that $q-1$ is not just a fitting parameter, but is directly related to the multiplicity fluctuations of the produced particles. For a  negative binomial distribution with parameter $k$, this simplifies to $q - 1 = 1/k$~\cite{Biro:2020kve}. So the power-law tail seen in Eq.~\eqref{eq:TsallisFit} is not an arbitrary feature of the fit, but arises naturally from the event-by-event fluctuations in multiplicity. A larger value of $q$ indicates a stronger power-law tail and a larger deviation from thermal-like behavior. This also corresponds to larger multiplicity fluctuations, and such enhanced fluctuations are, in turn, understood to arise from additional particle production via hard partonic (minijet) processes superimposed on the softer, bulk thermal emission. In jetty events, this hard tail is predominantly associated with jet fragmentation. In contrast, the limit $q\rightarrow1$ corresponds to an approximately Boltzmann--Gibbs-like exponential $p_{\rm T}$ spectrum. The Tsallis temperature $T_{\rm s}$ characterizes the effective energy scale of the low- and intermediate-$p_{\rm T}$ region, with larger $T_{\rm s}$ corresponding to a flatter exponential component. As discussed in Sec.~\ref{sec:tsallissec}, the events with large UE contributions are expected to exhibit relatively larger $T_{\rm s}$ together with smaller $q$, consistent with a spectrum dominated by soft particle production and close to the Boltzmann--Gibbs limit. In contrast, jet-dominated events generally exhibit larger $q$, reflecting the enhanced power-law tail from hard fragmentation, while the fitted $T_{\rm s}$ tends to be smaller because the hard component is primarily accommodated by the increase in $q$, and therefore $T_{\rm s}$ characterizes the remaining soft part of the spectrum. Therefore, studying both $q$ and $T_{\rm s}$ in different $\Delta\phi$ bins constitutes a two-dimensional spectral decomposition that probes the soft energy scale and the degree of non-extensivity. The $p_{\rm T}$ spectra in each $\Delta\phi$ bin were fitted using Eq.~\eqref{eq:TsallisFit} for different flattenicity and spherocity classes. The extracted parameters, $T_{\rm s}$ and $q$, are discussed below. The majority of the fits yield $\chi^2/\mathrm{NDF}$ values in the range of $0.3$--$4.0$, indicating an overall satisfactory description of the simulated $p_{\rm T}$ spectra.

Figure~\ref{fig:s0_qT_energy} shows the $\sqrt{s}$ dependence of $q$ (left) and $T_{\rm s}$ (right) as a function of $\Delta\phi$ for two transverse spherocity classes: jetty (upper) and isotropic (lower) pp collisions using PYTHIA8. Jetty events (upper panels, 0--20\% $S_0$) exhibit a higher $q$ and lower $T_{\rm s}$ compared to isotropic events (lower panels, 95--100\% $S_0$), at every collision energy and in every $\Delta\phi$ interval. Jetty events, dominated by fragmentation and parton showering, induce a prominent power-law tail at high $p_{\rm T}$, which is reflected as a higher value of $q$. Since the power-law component is primarily accommodated by an increase in $q$, therefore $T_{\rm s}$ characterizes the remaining low-$p_{\rm T}$ part of the spectrum, resulting in a relatively smaller value of $T_{\rm s}$. In contrast, the isotropic events are dominated by MPI-driven particle production leading to a $p_{\rm T}$ spectrum that is closer to an exponential shape which leads to a smaller value of $q$. The enhanced soft particle production from large $N_{\rm mpi}$ leads to a larger effective energy scale of the low- and intermediate-$p_{\rm T}$ region, reflected in higher $T_{\rm s}$ value.

The azimuthal dependence of $q$ and $T_{\rm s}$ differs between spherocity classes. In jetty events both quantities are strongly modulated with $\Delta\phi$, with maxima in the NS and AS regions, while at the lower collision energies a broad local maximum in $q$ also develops within the TS, which weakens with increasing $\sqrt{s}$. Following Eq.~\eqref{eq:q_multiplicity}, $q-1$ reflects the relative multiplicity fluctuations of the fitted sample rather than the hardness of the spectrum alone, so that a maximum of $q(\Delta\phi)$ marks the angular region of largest event-by-event fluctuation. In the NS and AS these fluctuations are intrinsic to jet fragmentation; as shown in Fig.~\ref{fig:mpi_selections} (Appendix~\ref{sec:app1}), they are governed by the leading semi-hard scattering rather than by the overall MPI activity. In the TS, the enhanced fluctuations may instead reflect the superposition of a relatively soft underlying component and a second component associated with jet fragmentation. This contribution can become more pronounced when the leading particle has low $p_{\rm T}$ and therefore provides a less accurate proxy for the underlying parton jet direction. At lower $\sqrt{s}$ the soft bulk thins out as $\langle N_{\rm mpi}\rangle$ decreases while the fragmentation cone widens, so the two contributions become comparable in yield and their combined fluctuation is largest, producing the local maximum. At $\sqrt{s}=13$~TeV the underlying event dominates and averages it out.

At first sight, the observed $\Delta\phi$ dependence of $T_{\rm s}$ in the jet-dominated regions may appear to contradict the event-class-dependent ordering, in which stronger jet dominance is associated with a smaller $T_{\rm s}$. However, these two comparisons are not equivalent. The event-shape class comparison is made between event samples with different $\langle N_{\rm ch}\rangle$ and $\langle\hat{p}_{\rm T}\rangle$ at the same $\Delta\phi$, where the enhanced hard component is reflected primarily in an increase of $q$, accompanied by a decrease in $T_{\rm s}$ through their anti-correlation. In contrast, the $\Delta\phi$-dependent comparison is performed within a fixed event class, where the NS and AS spectra develop a steeper soft component associated with jet fragmentation, resulting in a simultaneous increase of both $q$ and $T_{\rm s}$. Moreover, the hardening of the power-law tail of the $p_{\rm T}$ spectra with increasing collision energy is reflected in the corresponding increase of $q$~\cite{alice:prl2010,ALICE:vz}. In this event class, the $\sqrt{s}$ dependence of $T_{\rm s}$ is negligible, indicating that it is predominantly governed by the low- to intermediate-$p_{\rm T}$ component of jet fragmentation.

However, as one moves towards isotropic events, the azimuthal dependence of $q$ becomes mostly negligible across the TS and AS regions, with an increase appearing only near the NS boundary due to the trigger-particle selection bias. In contrast, $T_{\rm s}$ retains a mild but genuine modulation over the full $\Delta\phi$ range. It increases from the NS region and reaches a shallow maximum in the bins adjacent to the TS, then decreases towards a shallow minimum near the center of the AS region before increasing again across the second TS interval. This smooth, wave-like pattern is different from the sharp enhancement near the NS seen with the full markers, which is mainly a consequence of the trigger-particle bias rather than the underlying event topology. This behavior can be understood from the high-$S_0$ selection, which suppresses back-to-back dijet topologies and leaves particle production largely dominated by the UE. For isotropic events, the $\sqrt{s}$ dependence of both $q$ and $T_{\rm s}$ is also more pronounced. This does not contradict the observation that a larger UE contribution drives $q$ towards unity, since these two effects are compared in different ways. At a fixed $\sqrt{s}$, increasing the MPI activity adds more mutually uncorrelated soft production, reducing the relative multiplicity fluctuations and hence $q$. In contrast, when comparing different $\sqrt{s}$ values, both $\langle N_{\rm mpi}\rangle$ and the characteristic hard scale $\langle\hat{p}_{\rm T}\rangle$ increase (Fig.~\ref{fig:mpiPThatvssqrts}). The increase in the hard scale makes the power-law tail harder and consequently leads to a larger $q$. Thus, the UE does not necessarily move closer to the Boltzmann--Gibbs limit with increasing $\sqrt{s}$; rather, it moves away from it, refining the picture inferred at a single energy in Ref.~\cite{Mishra:2021hnr}.

Figure~\ref{fig:flat_qT_energy} presents the corresponding results for $1-\rho_{\rm ch}$ selected events. The overall behavior closely follows that observed for the $S_0$ selection. The jetty-like (0--20\%) $1-\rho_{\rm ch}$ class show larger $q$ and smaller $T_{\rm s}$ than the isotropic (95--100\%) $1-\rho_{\rm ch}$ class over the entire $\Delta\phi$ range and at all collision energies. Moreover, the weak increase of $q$ with $\sqrt{s}$ in isotropic events and the event-shape evolution of $T_{\rm s}$ are similar to $S_{0}$ selection. The azimuthal dependence also remains qualitatively unchanged: TS is characterized by lower $q$ than the NS, while the AS retains its topology-dependent behavior, exhibiting a pronounced enhancement in jetty events and a much weaker modulation in isotropic events. The smaller bump of $q$ in AS region is expected for flattenicity-based selection, in contrast to $S_0$, as $1-\rho_{\rm ch}$ has a broader coverage of $\hat{p}_{\rm T}$ than $S_0$. Additionally, despite $1-\rho_{\rm ch}$ being defined in the forward rapidity region ($2.2<\eta<5.0$), which is well separated from the midrapidity spectra used in the Tsallis fits, the same qualitative trends are observed in both $S_{0}$- and $1-\rho_{\rm ch}$-based event selections. This consistency disfavors an interpretation of the observed topology dependence as arising from autocorrelations between the classifier and the midrapidity spectra. It should be noted, however, that both classifiers retain a residual correlation with the charged-particle multiplicity. When integrated over $N_{\rm ch}$, the jetty classes also correspond to lower $\langle N_{\rm ch}\rangle$, while $q$ depends explicitly on $\langle n\rangle$, as shown in Eq.~\eqref{eq:q_multiplicity}.

The jetty-like flattenicity selection exhibits a different $\Delta\phi$ dependence of $T_{\rm s}$ than the $S_0$-based jetty events, which shows a $\sqrt{s}$ dependence. At the highest collision energies, distinct minima of $T_{\rm s}$ are observed in the TS and AS regions. As $T_{\rm s}$ quantifies the low and intermediate $p_{\rm T}$ shape, a weak $\Delta\phi$ dependence of $T_{\rm s}$ for the jetty events denotes the unbiased low UE and multi-jet topology, in contrast to $S_0$-based jetty events, which are prone to pick the dijet events leading to a peak in NS and AS regions. Moreover, as $\sqrt{s}$ decreases, $\Delta\phi$ dependence further weakens, and at the lowest energies $T_{\rm s}$ becomes nearly flat over the full azimuthal range while attaining larger values. This behavior indicates that the phase space for hard parton production is reduced at low collision energies, which leads to a decrease in the contrast between jet-dominated and UE-dominated regions. Consequently, particle production becomes increasingly governed by soft processes whose characteristic momentum scale varies only weakly with azimuthal angle. The larger overall value of $T_{\rm s}$ at low $\sqrt{s}$ should therefore not be interpreted as a harder jet spectrum. Rather, in the absence of a significant power-law component (smaller $q$), the Tsallis fit attributes a larger fraction of the spectral shape to the exponential component, resulting in a larger value of $T_{\rm s}$. At higher collision energies, where hard scattering and jet fragmentation become increasingly important, the power-law tail is predominantly accommodated through an increase in $q$, leading to a smaller fitted $T_{\rm s}$ and a prominent $\Delta\phi$ dependence.

\begin{figure*}[!t]
	\centering    
	\includegraphics[width=\textwidth]{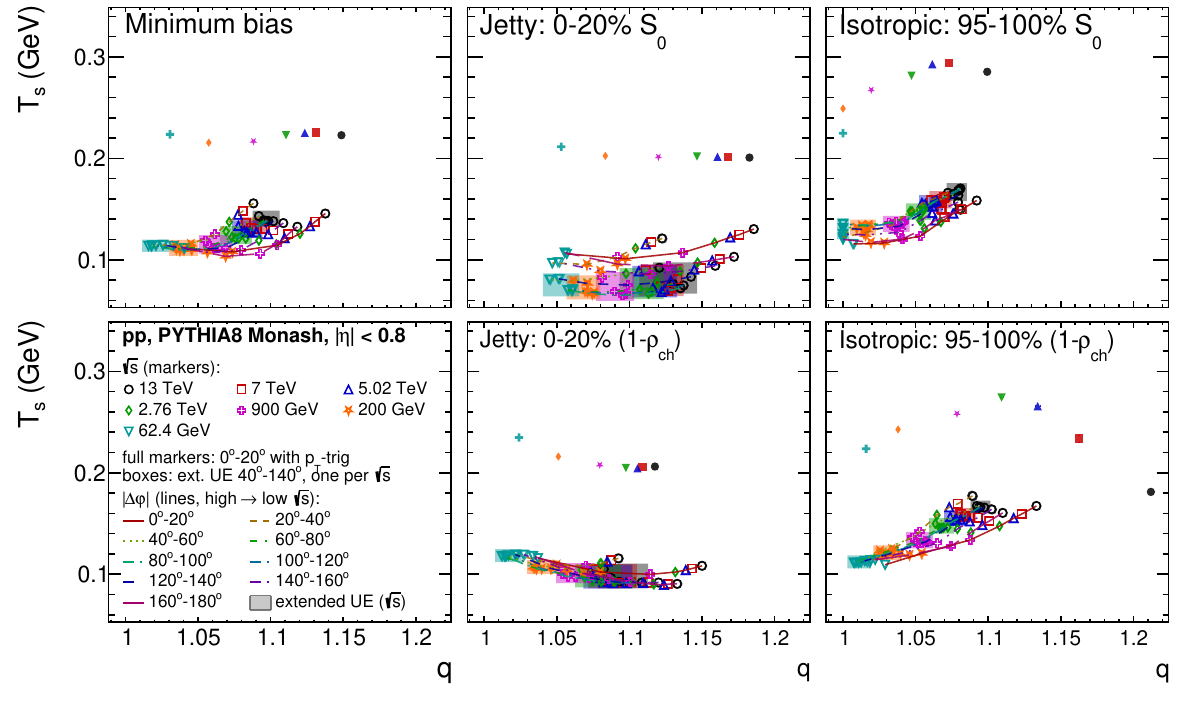}
    \vspace{-1cm}
	\caption{Tsallis thermometer representation ($T_{\rm s}$ versus $q$) for spherocity (upper) and flattenicity (lower) selected jetty (middle) and isotropic (right) event classes in different $\Delta\phi$ bins. The statistical uncertainties are smaller than the marker size.}
	\label{fig:tq_flat}
\end{figure*}

Interestingly, $q$ versus $\Delta\phi$  exhibit a shallow minimum immediately outside the near-side region. This feature persists across all collision energies, event-topology classes, and event-shape classifiers, indicating that it is not an artifact of the event selection. Furthermore, the position of the minimum shifts systematically with collision energy, moving from the transverse-side boundary at RHIC energies towards the near-side region at LHC energies. The same energy-dependent behavior is also observed in the events selected based on the hard-scattering scale ($\hat{p}_{\rm T}$) in Fig.~\ref{fig:mpi_selections} (see Appendix~\ref{sec:app1}), whereas it is considerably less pronounced in the events selected based on high $N_{\rm mpi}$. These observations suggest that the feature is primarily governed by the angular evolution of particle production correlated with the leading hard scattering. In contrast, $T_{\rm s}$ in the similar region exhibit only a weak azimuthal modulation, characterized by a broad enhancement around the same angular region, indicating that the energy-dependent shift is a feature of the non-extensive parameter $q$.

There is a related way to look at this: the AS peak itself becomes less pronounced as $\sqrt{s}$ decreases. At LHC energies, the AS enhancement of $q$ and $T_{\rm s}$ in Figs.~\ref{fig:s0_qT_energy} and \ref{fig:flat_qT_energy} is sharply localized around $\Delta\phi \approx \pi$. At lower energies, this peak broadens and its edges spread into the neighboring TS bins. The cause could be the same one discussed earlier: at lower $\langle \hat{p}_{\rm T} \rangle$, the leading particle becomes a less reliable tag for the true recoil-jet direction (Fig.~\ref{fig:mpiPThatvssqrts}(b)), so the AS fragmentation products become smeared across a wider angular range once the event is oriented relative to the leading particle rather than the true jet axis.  That the hard scattering itself is not actually weaker at low $\sqrt{s}$ is evident from Fig.~\ref{fig:mpi_selections} (Appendix~\ref{sec:app1}): when events are instead selected on the true hard-scattering scale $\hat{p}_{\rm T}$, the near-/away-side enhancement of $q$ is as strong, if not stronger, at lower $\sqrt{s}$. Since hard scatterings are rarer there, their effect stands out more clearly. Selections on high $N_{\rm mpi}$, on the other hand, remain comparatively flat at both energies, while low-$N_{\rm mpi}$ events reproduce the modulation seen for the hard-scattering selection. Thus, the fading AS peak observed in Figs.~\ref{fig:s0_qT_energy} and \ref{fig:flat_qT_energy} is a consequence of the leading particle becoming less well aligned with the primary hard-scattered jet axis at low $\sqrt{s}$, and not of a genuine weakening of the recoil. This is confirmed by the $p_{\rm T}^{\rm lead}>2$~GeV/$c$ check described above: with a genuine hard trigger the away-side enhancement of $q$ relative to the transverse region is comparable to, or larger than, its value at LHC energies.

It is worth emphasizing that in both Figs.~\ref{fig:s0_qT_energy} and~\ref{fig:flat_qT_energy}, the Tsallis parameters exhibit a flatter behaviour across the UE-dominated region, while showing a clear modulation in the NS and AS regions. This difference reflects the underlying physics. In the NS and AS, the spectra are still influenced by the hard partonic scattering and its associated fragmentation, so the Tsallis parameters retain some sensitivity to the jet axis and change with the angular distance from it. In the UE-dominated range, on the other hand, particle production is mainly driven by soft contributions from multi-parton interactions and beam-remnant activity, which are approximately uniform in azimuth. As a result, the extracted temperature and non-extensivity parameters show little dependence on $\Delta\phi$. This angular stability of $T_s$ and $q$ in the TS can therefore be regarded as a characteristic feature of the UE. Empirically, for TS region, the following relations is to be satisfied~\cite{Mishra:2021hnr}.
\begin{equation}
    \frac{\text{d} T_s}{\text{d} \Delta\phi} \neq 0 \quad \& \quad \frac{\text{d} q}{\text{d} \Delta\phi} \neq 0 \quad \text{for NS \& AS}
    \label{eq:NSASdiff}
\end{equation}
\begin{equation}
    \frac{\text{d} T_s}{\text{d} \Delta\phi} \approx 0 \quad \& \quad \frac{\text{d} q}{\text{d} \Delta\phi} \approx 0 \quad \text{for TS}
    \label{eq:TSdiff}
\end{equation}

The Eqs.~\eqref{eq:NSASdiff} and \eqref{eq:TSdiff} are most clearly satisfied for the isotropic events across collision energies and event-shape classes. However, for the jetty events, where the TS region receives contamination from jet-fragmentation and the misalignment between a jet axis and its leading particle, Eq.~\eqref{eq:TSdiff} does not hold true. 
\subsection{Tsallis thermometer}
\label{sec:tsallisthermo}
Figure~\ref{fig:tq_flat} presents the Tsallis thermometer representation, i.e., the correlation between the Tsallis temperature $T_{\rm s}$ and the non-extensivity parameter $q$, with every point corresponding to a $\Delta\phi$ bin at a specific collision energy, for the spherocity-based (upper row) and flattenicity-based (lower row) event selection, respectively. In the minimum-bias sample (upper left panel), the $\Delta\phi$ bins for a given $\sqrt{s}$ form a compact, $\sqrt{s}$-dependent group rather than a single continuous trajectory. Here the seven $\sqrt{s}$ groups are arranged one above another primarily along the $T_{\rm s}$ axis, from $T_{\rm s}\approx0.11$~GeV at $62.4$~GeV to $T_{\rm s}\approx0.14$--$0.16$~GeV at $13$~TeV, with a comparatively modest spread in $q$ within each $\sqrt{s}$ group coming from the different $\Delta\phi$ slices. The full markers in Fig.~\ref{fig:tq_flat} denote the $0^\circ$--$20^\circ$ bin evaluated with the trigger particle included. Since the leading particle enters this bin once per event and is by construction the hardest particle. These points lie well above the bulk at $T_{\rm s}\approx0.2$--$0.3$~GeV for all $\sqrt{s}$ and event classes, which is approximately 100 MeV higher than the rest without the leading particle. They also do not follow the $\sqrt{s}$ ordering observed for the other bins. The corresponding $(q,,T_{\rm s})$ pairs reflect the selection effect rather than a thermodynamic property of the source.

Moreover, the event-shape-dependent study reveals that the isotropic class (right column) produces the cleanest thermometer correlation. Here, the energy groups line up along a well-defined, monotonically rising diagonal in the $(q,\,T_{\rm s})$ plane, from the lowest $\sqrt{s}$ points near $(q,\,T_{\rm s})\approx(1.00,\,0.12~\text{GeV})$ up to the highest $\sqrt{s}$ points near $(1.09,\,0.17~\text{GeV})$ for $S_0$ selection, and an even more extended range, up to $T_{\rm s}\approx0.20$~GeV, for $1-\rho_{\rm ch}$ selection. In contrast, the jetty class (middle column) is the most complex of the three event shape cases. Here, the different $\sqrt{s}$ groups overlap heavily in both $q$ and $T_{\rm s}$ rather than separating into distinct bands, so no single well-defined trajectory can be observed. This is consistent with the picture discussed in Sec.~\ref{sec:delphivsTandq}, in which the jetty selection is precisely the class where the hard power-law tail and the low-$p_{\rm T}$ thermal component compete most strongly for the same spectral shape, bin by bin and energy by energy, producing scatter in the fitted $(q,\,T_{\rm s})$ pair rather than a clean correlation. It should be noted that the elongated shape of these $\sqrt{s}$ dependent groups partially reflects the intrinsic anti-correlation of $q$ and $T_{s}$ discussed in Sec.~\ref{sec:tsallissec}.

Comparing the two classifiers within the jetty class, the spherocity selection spans the widest range in $q$, while the flattenicity selection is more compressed, with $T_{\rm s}$ confined to a narrow, nearly flat band that even dips slightly at intermediate $q$ before rising again at the highest values. This compressed, weakly-structured jetty band for $1-\rho_{\rm ch}$ is consistent with the near-flat, only gradually developing $\Delta\phi$ dependence of $T_{\rm s}$ already reported for flattenicity-selected jetty events in Sec.~\ref{sec:delphivsTandq}, in contrast to the more strongly modulated spherocity-selected jetty events. The isotropic panels of both rows, meanwhile, retain the same $\sqrt{s}$-dependent rising trend, with flattenicity extending to higher $T_{\rm s}$ than spherocity at comparable $q$. To conclude, Fig.~\ref{fig:tq_flat} shows that the well-organized $q$--$T_{\rm s}$ correlation associated with the Tsallis thermometer is a feature of the UE-dominated, isotropic event class, while jetty event selection degrades rather than sharpens this correlation, and does so somewhat more for spherocity than for flattenicity.

A particularly informative feature of Fig.~\ref{fig:tq_flat} is the behaviour of the $\Delta\phi$ bins belonging to the extended UE region, $40^\circ<|\Delta\phi|<140^\circ$, indicated by the colored boxes drawn separately for each $\sqrt{s}$. These bins do not spread out along the trajectory formed by the full set of sliding-angle bins. Instead, they group together in a small area of the $(q,\,T_{\rm s})$ plane. This area is much smaller than the range covered by the NS and AS bins at the same energy. All bins inside the extended UE window are therefore described by nearly the same $(q,\,T_{\rm s})$ pair. This implies that the extended UE window is not simply an average over different angular regions, but behaves like a single, angularly uniform source. The grouping is the $(q,\,T_{\rm s})$-plane counterpart of the small angular derivatives given in Eq.~\eqref{eq:TSdiff}. It is also a stronger statement, since $T_{\rm s}$ and $q$ could individually vary little while still compensating each other bin by bin, whereas the grouping requires both to be stable at the same time.

For the minimum-bias and isotropic classes the group stays equally compact from $\sqrt{s}=62.4$~GeV to $13$~TeV, while its position shifts. It moves upward along the $T_{\rm s}$ axis in the same way as the minimum-bias groups discussed above, together with a smaller increase in $q$. Within these classes the compactness of the extended UE region is therefore universal, while its location is not. The UE follows its own rising trajectory, roughly parallel to the isotropic-class diagonal but shifted away from it. Two points follow. First, the pair $(q_{\rm UE},\,T_{\rm s}^{\rm UE})$ gives a simple energy-dependent reference point for the soft, MPI-dominated part of pp collisions, which can be compared with the same quantity in larger systems. Second, how close the UE is to the Boltzmann--Gibbs limit depends on the collision energy. At every $\sqrt{s}$ the extended UE group has the lowest $q$ of all angular regions, as in Ref.~\cite{Mishra:2021hnr}, but it moves away from $q\rightarrow1$ as $\sqrt{s}$ increases. The description of the UE as nearly Boltzmann--Gibbs-like is thus specific to LHC energies. At RHIC energies the UE is closer to this limit, and the reason is the smaller hard scale rather than a larger UE activity, in line with Sec.~\ref{sec:delphivsTandq}.  

The jetty class is the exception to this behaviour. Here the boxes are elongated rather than compact, and they overlap between neighbouring collision energies instead of staying separated in $\sqrt{s}$. The energy trend described above therefore cannot be followed in this class. The reason is the same one described in Sec.~\ref{sec:delphivsTandq}. In jetty events the extended UE window is contaminated by fragmentation products at every collision energy, so Eq.~\eqref{eq:TSdiff} no longer holds and the bins no longer group together. The contamination is diluted by the underlying event at the highest energies and becomes relatively more important at RHIC energies, where the fragmentation cone is wider. The two causes leave different signatures: the elongation at the highest energies is a property of the jetty selection itself, while the elongation at RHIC energies is driven by the leading-particle tagging and is reduced when a hard trigger is required. How tightly the extended UE box is grouped can therefore be used as a simple measure of how well a given event-shape selection isolates the underlying event, and the contrast between the isotropic and jetty classes shows that the grouping is a property of UE-dominated events rather than of the angular window alone.

\section{Summary and outlook}
\label{summary}

In summary, we have presented a multi-differential study of charged-particle production in pp collisions over a broad collision-energy range, from $\sqrt{s}=62.4$~GeV to $13$~TeV, using PYTHIA8 with transverse spherocity and charged-particle flattenicity as event-shape classifiers. The main findings are given as follows.

\begin{itemize}

\item The azimuthal structure of particle production exhibits a universal topology from RHIC to LHC energies, characterized by enhanced particle production in the NS and AS regions and a comparatively flat UE dominated TS region. Although the overall particle production increases with collision energy due to enhanced $N_{\rm mpi}$, the relative angular structure remains unchanged with a change in $\sqrt{s}$.

\item The combined study of normalized charged-particle multiplicity and transverse spherocity shows that the UE region can be extended to $40^\circ < |\Delta\phi| < 140^\circ$ (or equivalently $2\pi/9 \le |\Delta\phi| < 7\pi/9$). This extended UE region preserves the characteristic soft-QCD dominated particle-production across $\sqrt{s}$ regions, providing a larger azimuthal acceptance for UE studies.

\item The Tsallis parameters reveal a clear dependence on event topology. Jetty events exhibit larger non-extensivity parameter $q$ and lower Tsallis temperature $T_{\rm s}$ compared to isotropic events for all $\sqrt{s}$ and $\Delta\phi$ regions. The enhanced $q$ values in the NS and AS regions are primarily driven by hard-QCD processes rather than by the overall increase in MPI activity.

\item Transverse spherocity separates the jetty and isotropic classes more strongly in the non-extensivity parameter $q$, whereas charged-particle flattenicity yields a cleaner and more extended $q$--$T_{\rm s}$ correlation; the two classifiers are thus complementary probes of the event topology.

\item On the Tsallis thermometer, the $\Delta\phi$ bins belonging to the extended UE region group into a compact cluster in the $(q,\,T_{\rm s})$ plane. For minimum-bias and isotropic events this cluster stays equally compact from $\sqrt{s}=62.4$~GeV to $13$~TeV while its position shifts towards higher $T_{\rm s}$ and, to a smaller extent, higher $q$. The compactness of the extended UE region is therefore universal, but its location is energy dependent, and the UE moves away from the Boltzmann--Gibbs limit as $\sqrt{s}$ increases. In jetty events the cluster is elongated and the energy groups overlap, so how tightly the extended UE bins group together also measures how well an event-shape selection isolates the underlying event.

\end{itemize}

These results establish the combination of azimuthal differential measurements, event-shape classification, and Tsallis phenomenology as a powerful approach to disentangle the contributions of hard scatterings, MPIs, and soft-QCD dynamics in pp collisions. Comparisons with experimental measurements and extensions to identified particles, heavy-flavour hadrons, and other small collision systems will provide further constraints on the microscopic origin of non-extensive particle production. A systematic comparison of similar measurements from different experiments would shed light on the interplay of hard and soft-QCD processes and their evolution with collision energy.

The present results provide a baseline for investigating the evolution of particle-production mechanisms with increasing collision-system size. Extending this multi-differential framework to larger collision systems, including OO, Ne--Ne, and Pb--Pb collisions, will enable a systematic study of how initial-state geometry, and competing effects from hard and soft-QCD processes evolve from small to large collision systems. Particularly, comparing the energy and system-size dependence of event-shape observables, UE characteristics, and Tsallis thermodynamic parameters can provide insight into the transition from MPI-dominated particle production in pp collisions to systems where nuclear geometry effects, along with signatures often associated with collective-like behavior, become increasingly relevant. Such studies will help establish whether the observed non-extensive behavior and topology-dependent particle-production patterns represent universal features of QCD particle production or emerge from system-size-dependent dynamics.

\section*{Acknowledgements}
The authors would like to thank the Hungarian National Research, Development and Innovation Office (NKFIH) under Contract No. NKFIH ADVANCED\_25 K153456, NKFIH NEMZ\_KI-2022-00058, 2024-1.2.5-TET-2024-00022, and Wigner Scientific Computing Laboratory (WSCLAB, the former Wigner GPU Laboratory). Mexican Conacyt project CF2042 and the support of the Coordinacion de la Investigacion Cientifica, Mexico. Discussions on the fit procedure with G.~B\'\i r\'o and T.S.~Bir\'o are highly appreciated.


\appendix

\section{$\sqrt{s}$ versus ${\rm d}N_{\rm ch}/{\rm d}\eta$}
\label{sec:sqrtsvsnch}
\begin{figure}[!h]
    \centering
    \includegraphics[width=0.95\linewidth]{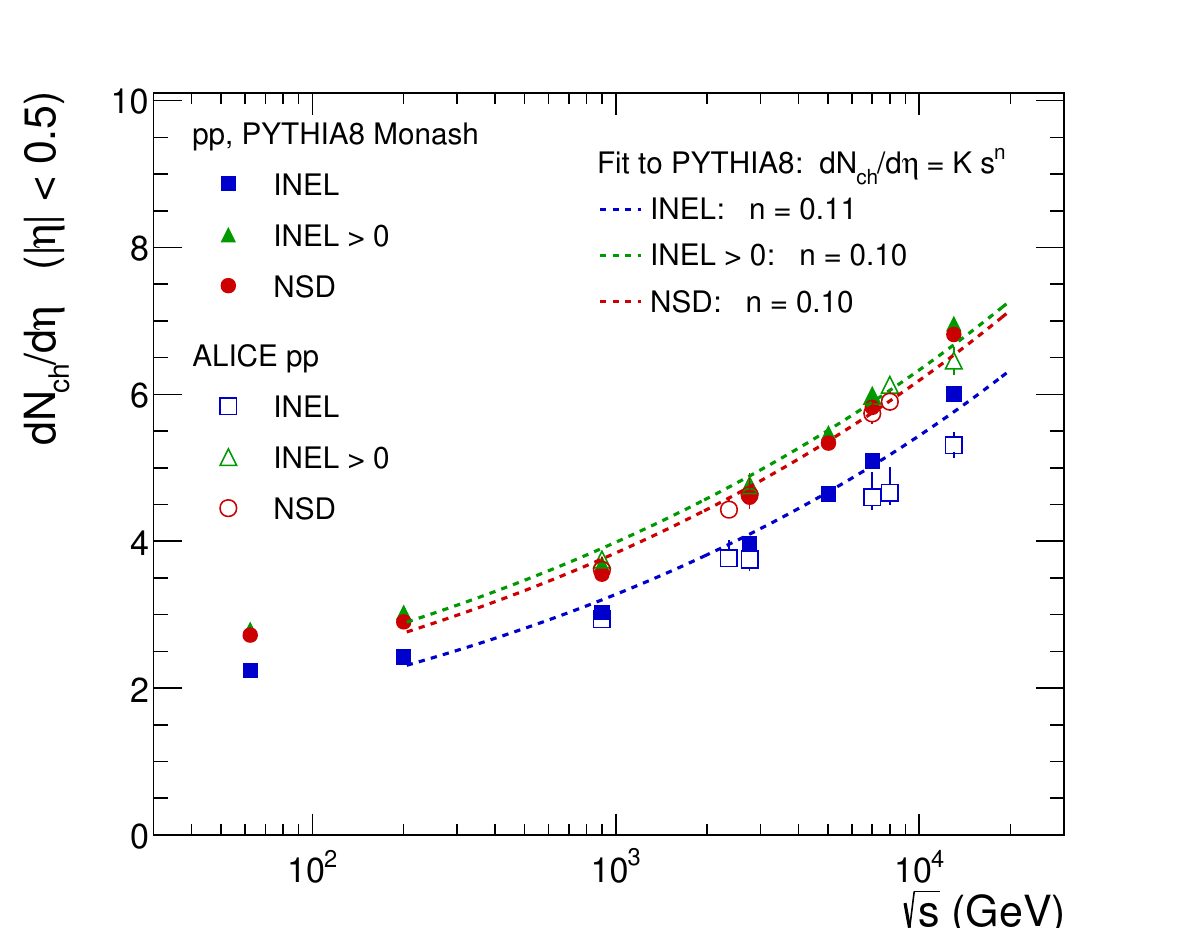}
    \caption{${\rm d}N_{\rm ch}/{\rm d}\eta$ at $|\eta| < 0.5$ versus $\sqrt{s}$ in pp collisions, for the INEL (squares), INEL${}>0$ (triangles) and NSD (circles) event classes. Filled symbols: PYTHIA~8 Monash; open symbols: ALICE~\cite{ALICE:2015olq,ALICE:2010cin,ALICE:2015qqj}. Dashed lines are power-law fits $K\,s^{n}$ to the PYTHIA points, with $n$ given in the legend. The statistical uncertainties of the PYTHIA8 results are smaller than the marker size.}
    \label{fig:sqrtsvsnch}
\end{figure}

Figure~\ref{fig:sqrtsvsnch} shows the energy dependence of the charged-particle pseudorapidity density, $\mathrm{d}N_{\mathrm{ch}}/\mathrm{d}\eta$, measured at midrapidity ($|\eta| < 0.5$) in proton--proton collisions over a centre-of-mass energy range spanning roughly $\sqrt{s} = 62.4$~GeV to $13$~TeV. Three event classes are considered: inelastic (INEL), inelastic events with at least one charged particle produced in $|\eta| < 1$ (INEL$>0$), and non-single-diffractive (NSD) collisions. For each class, predictions from the PYTHIA8 event generator using the Monash tune are compared directly to the corresponding ALICE pp measurements~\cite{ALICE:2015olq,ALICE:2010cin,ALICE:2015qqj}. 

PYTHIA8 reproduces the overall trend of the ALICE data reasonably well, with the multiplicity density rising smoothly and monotonically with $\sqrt{s}$ for all three event classes. To quantify the energy dependence, the PYTHIA8 predictions are fitted with a power-law function of the form $\mathrm{d}N_{\mathrm{ch}}/\mathrm{d}\eta = K\,s^{\,n}$, where $K$ is a normalization constant and $n$ characterizes the rate of growth with collision energy. The power-law fit to the PYTHIA8 points leads to $n=0.11$ for INEL, $n=0.1$ for both INEL$>0$ and NSD events.

\section{$\sqrt{s}$ dependence of $\langle N_{\rm mpi}\rangle$ and $\langle\hat{p}_{\rm T}\rangle$}
\label{sec:energydepofmpi}

\begin{figure}
    \centering
    \includegraphics[width=0.9\linewidth]{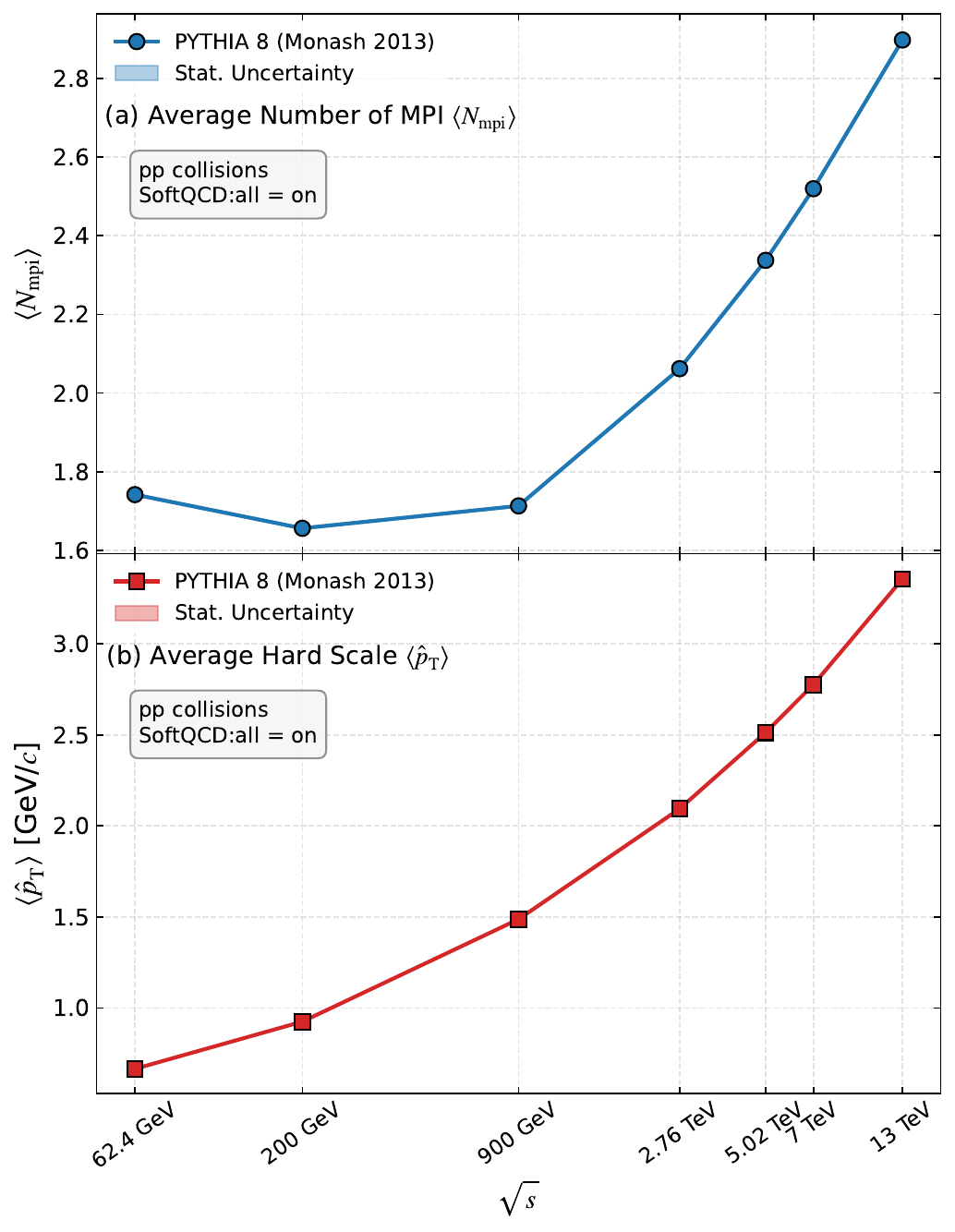}
    \caption{Collision energy dependence of (a) $\langle N_{\rm mpi}\rangle$ and (b) $\langle\hat{p}_{\rm T}\rangle$  in pp collisions using PYTHIA8. The statistical uncertainties are smaller than the marker size.}
    \label{fig:mpiPThatvssqrts}
\end{figure}

\begin{figure*}
    \centering
    \includegraphics[width=0.38\linewidth]{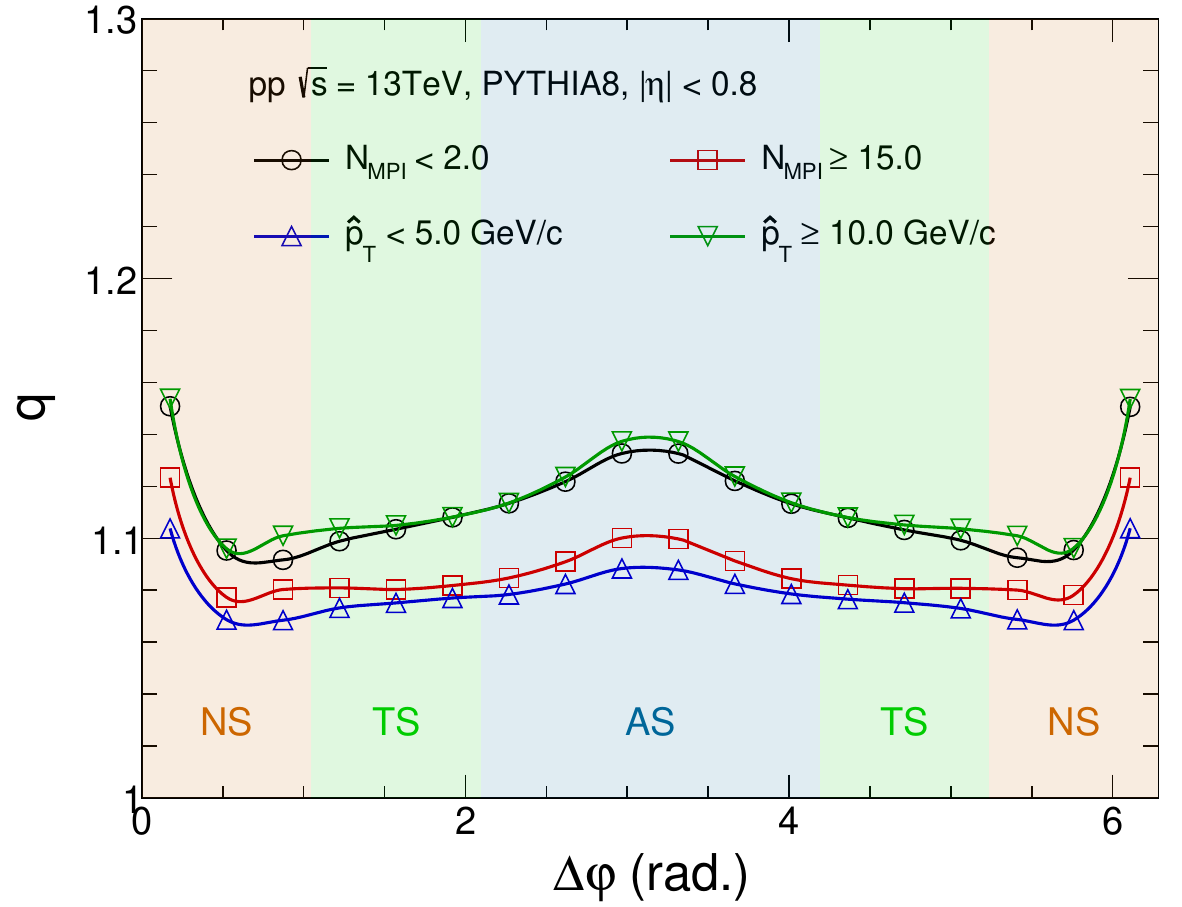}
    \includegraphics[width=0.38\linewidth]{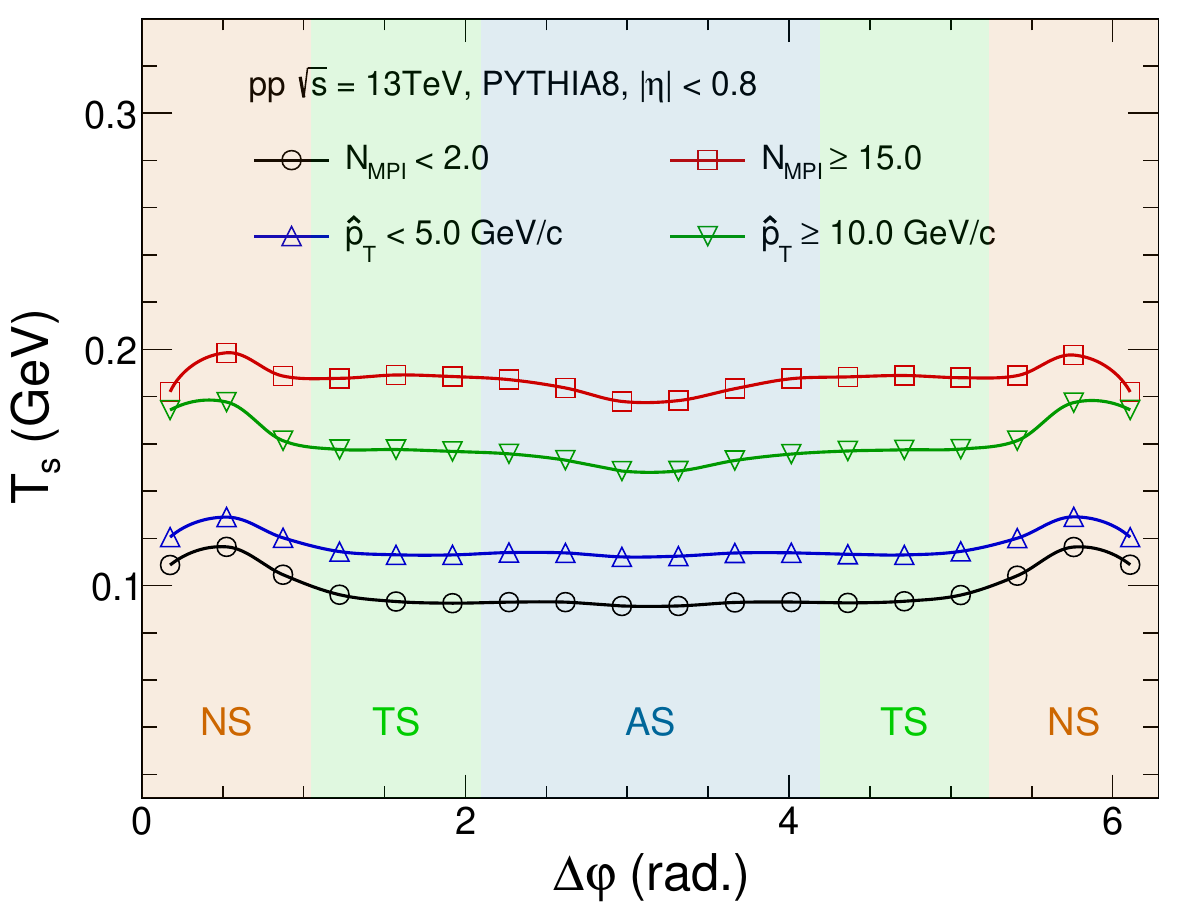}
    \includegraphics[width=0.38\linewidth]{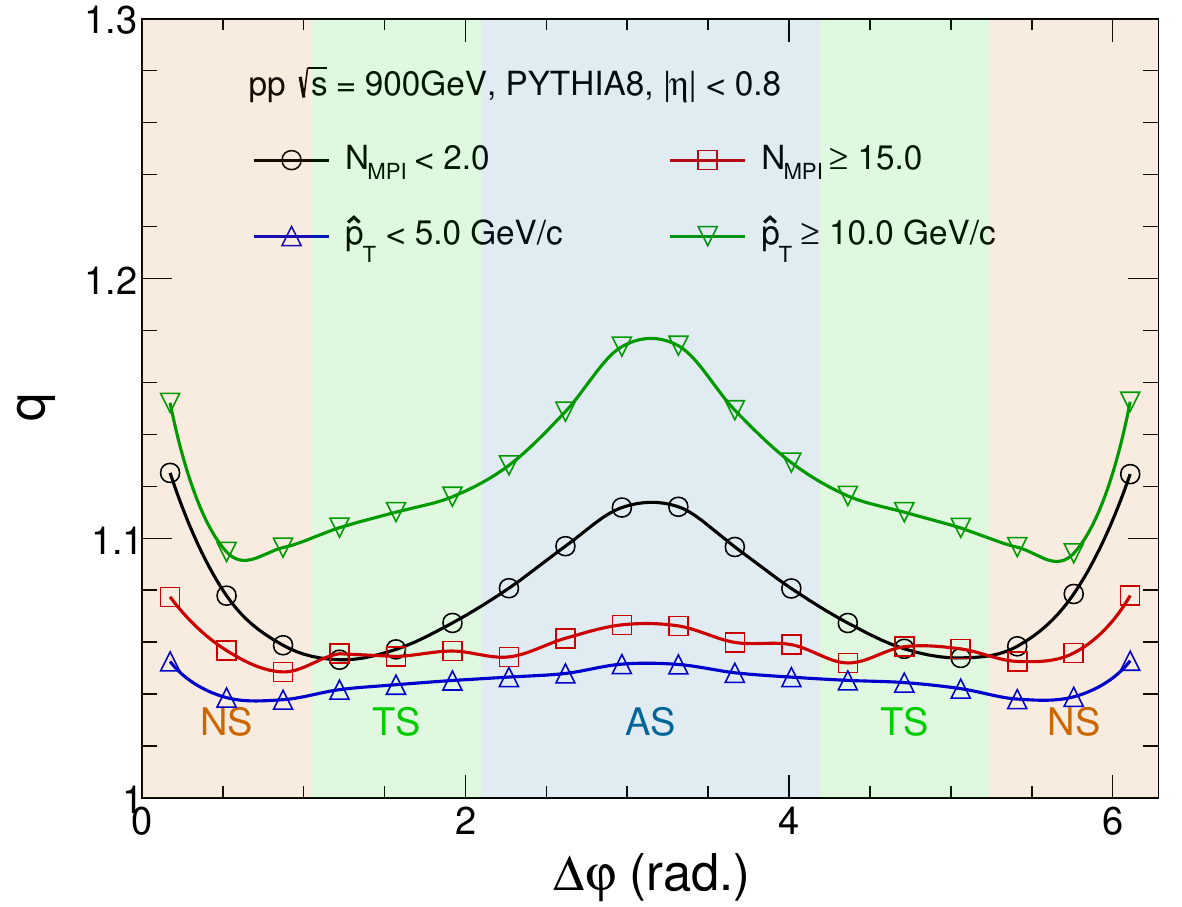}
    \includegraphics[width=0.38\linewidth]{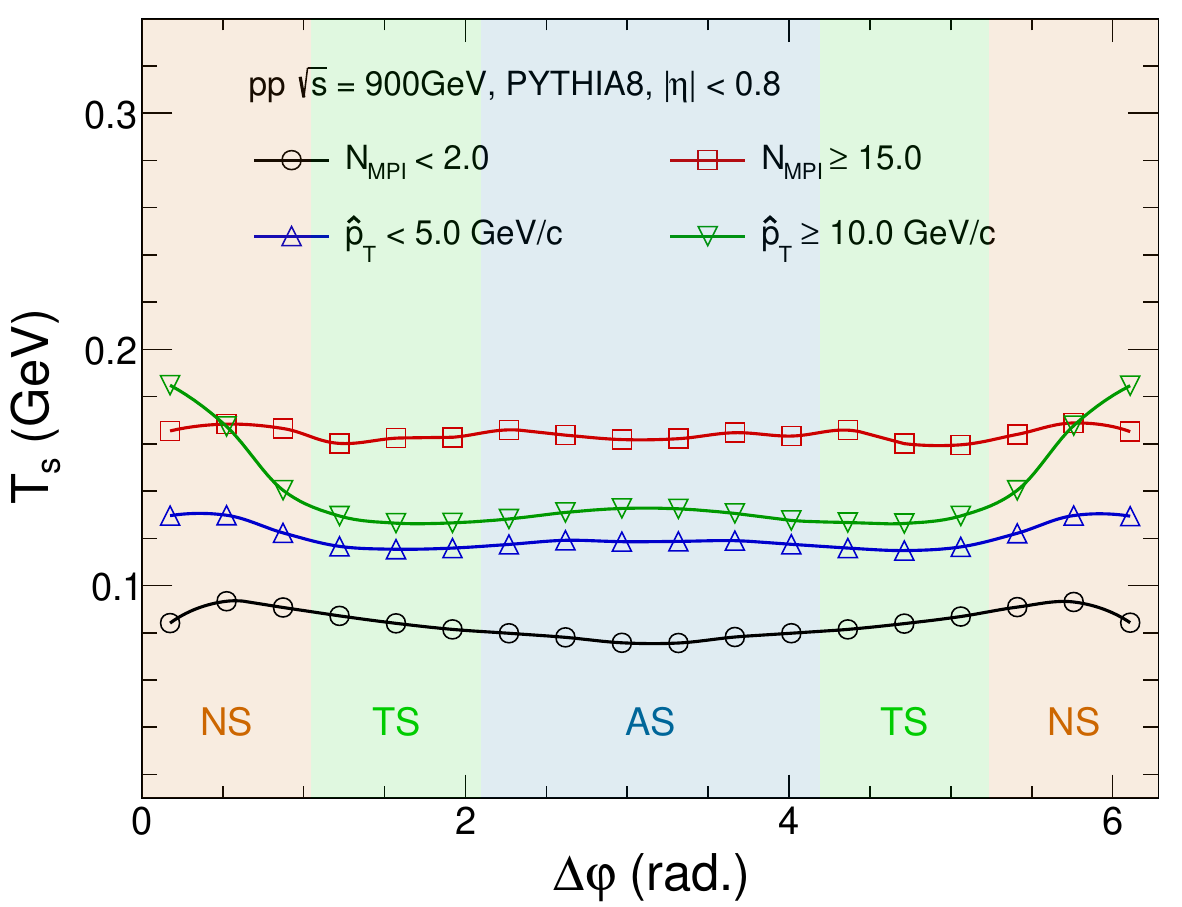}
	\caption{Comparison of Tsallis parameters, $q$ (left) and $T_{\rm s}$ (right) for event selected based on $N_{\rm mpi}$ and $\hat{p}_{\rm T}$ at $\sqrt{s}=13$ TeV (upper) and 900 GeV (lower). The statistical uncertainties are smaller than the marker size.}
	\label{fig:mpi_selections}
\end{figure*}

Figure~\ref{fig:mpiPThatvssqrts} shows the $\sqrt{s}$ dependence of the average number of multi-parton interactions, $\langle N_{\mathrm{mpi}} \rangle$, and the average hard-scattering scale, $\langle \hat{p}_{\mathrm{T}} \rangle$, in minimum-bias pp collisions generated with PYTHIA~8 using the Monash tune. Panel~(a) shows that $\langle N_{\mathrm{mpi}} \rangle$ exhibits a modest variation between RHIC energies and $\sqrt{s}=900$~GeV, followed by a pronounced increase toward LHC energies, indicating a substantial enhancement in the probability of multiple partonic interactions at higher collision energies. Panel~(b) shows that a monotonic increase of $\langle \hat{p}_{\mathrm{T}} \rangle$ with increasing $\sqrt{s}$, reflecting the growing contribution of harder partonic scatterings. Together, these observables indicate that increasing collision energy enhances both the underlying event activity through additional multi-parton interactions in PYTHIA and the characteristic momentum scale of the produced partons.

\section{Role of $N_{\rm mpi}$ and $\hat{p}_{\rm T}$}
\label{sec:app1}
To isolate the role of MPI from that of the leading hard process directly, Fig.~\ref{fig:mpi_selections} compares the $\Delta\phi$ dependence of $q$ (left) and $T_{\rm s}$ (right) for events selected on the number of MPI, $N_{\rm mpi}$, and on the transverse momentum of the hardest parton-level scattering, $\hat{p}_{\rm T}$, at $\sqrt{s}=13$~TeV (upper) and $900$~GeV (lower). At $\sqrt{s}=13$~TeV, the low-$N_{\rm mpi}$ and hard-scattering ($\hat{p}_{\rm T}\geq 10$ GeV/$c$) event selections exhibit nearly identical $q(\Delta\phi)$ behavior, with a pronounced enhancement in the away-side region, whereas the high-$N_{\rm mpi}$ and soft-scattering ($\hat{p}_{\rm T}< 5$ GeV/$c$) event selections show a relatively flatter dependence. This demonstrates that the near- and away-side enhancement of $q$ is primarily driven by the presence of a hard scattering rather than by the overall level of multiparton activity. Selecting events with a hard leading scattering reproduces the modulation observed for low-$N_{\rm mpi}$ events, while selecting events with large $N_{\rm mpi}$ substantially reduces it, despite the larger underlying-event activity. In contrast, the ordering of the $T_{\rm s}$ distributions is different. Here, high-$N_{\rm mpi}$ events consistently exhibit the largest values across $\Delta\phi$ regions, the hard-scattering selection lies at intermediate values with only mild azimuthal variation, and the low-$N_{\rm mpi}$ and soft-scattering selections remain lower with relatively weak modulation. At $\sqrt{s}=900$~GeV, the same qualitative pairing between the event selections is retained, although the $\Delta\phi$ dependence of $q$ becomes relatively stronger. The hard-scattering selection shows the most pronounced away-side enhancement, the low-$N_{\rm mpi}$ selection follows the same trend with a reduced amplitude, while the high-$N_{\rm mpi}$ and soft-scattering selections remain comparatively flat with a much larger variation between the near-, transverse-, and away-side regions than observed at $13$~TeV. The corresponding $T_{\rm s}$ distributions preserve the same ordering, although the separation between the high-$N_{\rm mpi}$ and hard-scattering selections is reduced compared with the higher collision energy. Taken together, the results at both energies demonstrate that the characteristic near-/away-side enhancement of $q$, and its evolution with $\sqrt{s}$ discussed in Sec.~\ref{sec:delphivsTandq}, is governed predominantly by the leading hard scattering. This signature is more distinctive at $\sqrt{s}=900$~GeV, where hard scatterings are comparatively rare, whereas at $\sqrt{s}=13$~TeV they are much more common and therefore less discriminative at higher energies.
\end{document}